\documentclass[twocolumn,showpacs,preprintnumbers,amsmath,amssymb]{revtex4}
\usepackage[dvips]{graphicx}
\def\gsim{\buildrel {\textstyle >}\over {_\sim}}
\def\lsim{\buildrel {\textstyle <}\over {_\sim}}

\begin{document}
\title{
Monte Carlo studies of the chiral and spin orderings of the three-dimensional Heisenberg spin glass
}
\author{Dao Xuan Viet and Hikaru Kawamura}
\affiliation{Department of Earth and Space Science, Faculty of Science,
Osaka University, Toyonaka 560-0043,
Japan}
\begin{abstract}
 The nature of the ordering of the three-dimensional isotropic Heisenberg spin glass with nearest-neighbor random Gaussian coupling is studied by extensive Monte Carlo simulations. Several independent physical quantities are measured both for the spin and for the chirality, including the correlation-length ratio, the Binder ratio, the glass order parameter, the overlap distribution function and the non-self-averageness parameter. By controlling the effect of the correction-to-scaling, we have obtained a numerical evidence for the occurrence of successive chiral-glass and spin-glass transitions at nonzero temperatures, $ T_{CG} > T_{SG} > 0$. Hence, the spin and the chirality are decoupled in the ordering of the model. The chiral-glass exponents are estimated to be $\nu_{CG}=1.4\pm 0.2$ and $\eta_{CG}=0.6\pm 0.2$, indicating that the chiral-glass transition lies in a universality class different from that of the Ising spin glass.  The possibility that the spin and chiral sectors undergo  a simultaneous Kosterlitz-Thouless-type transition is ruled out. The chiral-glass state turns out to be non-self-averaging, possibly accompanying a one-step-like peculiar replica-symmetry breaking. Implications to the chirality scenario of experimental spin-glass transitions are discussed.
\end{abstract}
\maketitle
\section{Introduction}
 Spin glasses (SGs) are the type of random magnets possessing both ferromagnetic and antiferromagnetic couplings, and are characterized by frustration and randomness. The ordering of SG has been studied quite extensively as a typical example of ``complex system'' and continues to give an impact on surrounding areas \cite {review}. Most of theoretical and numerical studies on the SG ordering have been based on a simplified statistical model called the Edwards-Anderson (EA) model, in which the spins are put on each site of a regular lattice and interact via the random coupling taking both positive (ferromagnetic) and negative (antiferromagnetic) signs \cite{EA}. For the {\it Ising\/} EA model in three dimensions (3D), it is now well established that the model exhibits an equilibrium SG transition at a finite temperature \cite{Ogielski85,BhattYoung85,Ballesteros00,Katzgraber06,Jorg06,Campbell06,Hasenbusch08}. The critical exponents of the SG transition evaluated by Monte Carlo (MC) simulations were compared favorably with those determined experimentally for the Ising-like SG compound FeMnTiO$_{3}$ \cite{Gunnarsson91}. 

 Meanwhile, many of real SG magnets, including the well-studied canonical SGs, {\it i.e.\/}, dilute magnetic alloys such as  CuMn, AuFe and AgMn, are the {\it Heisenberg-like\/} magnets possessing only weak magnetic anisotropy. Thus, an isotropic Heisenberg EA model, rather than the strongly anisotropic Ising EA model, is expected to be a more realistic model. Experimentally, the existence of a finite-temperature SG  transition and of a thermodynamic SG state  in real Heisenberg-like SG material has been established \cite{review}. 

 For the 3D isotropic Heisenberg EA model, earlier numerical studies suggested, in apparent contrast to the experimental observation, that the model exhibited only a zero-temperature transition \cite{Banavar82,McMillan85,Olive86,Matsubara91,Yoshino93}. Meanwhile, one of present author (H.K.) suggested that the model might exhibit a finite-temperature transition {\it in its chiral sector\/} \cite{Kawamura92}. Chirality is a multispin variable representing the handedness of the noncollinear or noncoplanar structures induced by frustration. It has subsequently been suggested that, in the ordering of the 3D Heisenberg SG, the chirality was ``decoupled'' from the spin, the chiral-glass (CG) order taking place at a temperature higher than the SG order, {\it i.e.\/}, $T_{CG} > T_{SG}$ \cite{Kawamura98,HukuKawa00,HukuKawa05,Kawamura07,Kawamura09}. Based on such a spin-chirality decoupling picture of the 3D isotropic Heisenberg SG, a chirality scenario of experimental SG transition was proposed \cite{Kawamura92,Kawamura96,Kawamura07,Kawamura09}: According to this scenario, the chirality is a hidden order parameter of SG transition. Real SG transition of weakly anisotropic SG magnets is then a ``disguised'' CG transition, where the chirality is mixed into the spin sector via a weak random magnetic anisotropy. For a recent review, the reader is referred to Ref.\cite{Kawamura09}

 The chirality scenario is capable of explaining several long-standing puzzles concerning the experimental SG transition \cite{review} in a natural way, such as the origin of the non-Ising critical exponents observed in canonical SGs \cite{Kawamura07,Kawamura09}, the apparent absence of the Heisenberg-to-Ising crossover in the measured nonlinear susceptibility \cite{Kawamura07,Kawamura09}, and the origin of the mean-field-like transition lines (the so-called AT and GT lines) often observed experimentally in the SG phase diagram in magnetic fields \cite {KawaIma01,ImaKawa04,Kawamura07,Kawamura09,Campbell99}. The chirality scenario remains to be an attractive hypothesis in consistently explaining various experimental observations for canonical SGs, and hence, it is an important task to examine the validity of the basic assumption underlying this scenario, {\it i.e.\/}, the occurrence of the spin-chirality decoupling in the 3D isotropic Heisenberg SG.

 In recent numerical studies of the 3D Heisenberg EA model, consensus now seems to appear that the 3D Heisenberg SG indeed exhibits a finite-temperature transition \cite{Kawamura92,Kawamura95,Kawamura98,HukuKawa00,HukuKawa05,Kawamura96,Kawamura07,Kawamura09,Matsubara00,Endoh01,Matsubara01,Nakamura02,Matsumoto02,LeeYoung03,BerthierYoung04,Picco05,Campos06,LeeYoung07}. However, the nature of the transition, especially whether the model really exhibits the spin-chirality decoupling, is still under hot debate \cite{Matsubara00,Matsubara01,Endoh01,Nakamura02,LeeYoung03,BerthierYoung04,Picco05,Campos06,LeeYoung07}. The present situation is not completely satisfactory.  Mentioning some of the recent numerical works: Hukushima and Kawamura studied the model with the random $\pm J$ coupling of modest lattice sizes $L\leq 20$  ($L$ being the linear dimension) but with a rather small number of samples of $N_{s}=32$ (for their largest $L$) \cite{HukuKawa05}. These authors then concluded that $T_{CG}$ is certainly finite, while $T_{SG}$ is either zero or nonzero but less than $T_{CG}$, {\it i.e.\/}, $T_{SG} < T_{CG}$, supporting the spin-chirality decoupling scenario. By contrast, Lee and Young claimed on the basis of their data of the correlation-length ratio $\xi/L$ of the model with the Gaussian coupling that the spin and chirality order at a common temperature, thus no spin-chirality decoupling \cite{LeeYoung03,LeeYoung07}. However, their data suffer from either small lattice sizes of only $L\leq 12$ \cite {LeeYoung03} or small number of samples of $N_{s}=56$ \cite {LeeYoung07}. Recently, Campos {\it et al\/} simulated the same model to much larger lattices $L=32$ with larger number of samples $N_{s}=1000$, but no data below the transition temperature \cite {Campos06}. Campos {\it et al\/} claimed that the chiral and spin sectors undergo simultaneously a Kosterlitz-Thouless (KT) transition with massive logarithmic corrections. This interpretation, however, was criticized in Ref. \cite {Campbell07} .
 
 Under such circumstances, we perform here a large-scale MC simulation of the 3D Heisenberg SG in order to shed further light on the nature of its spin and chirality ordering. We exceed the previous simulations by simulating the system as large as $L=32$ to temperatures considerably lower than $T_g$ for large number of samples of order $N_s\simeq 10^3$. Note that none of the previous simulations satisfied all these criteria simultaneously. More importantly, we calculate several independent physical quantities including the correlation-length ratios, the Binder ratios, the glass order parameters,  the overlap distribution functions and the non-self-averageness parameters, trying to draw consistent picture from these independent quantities, whereas Refs.\cite{LeeYoung03,Campos06,LeeYoung07} concentrated almost exclusively on the correlation-length ratio. By controlling the correction-to-scaling effect in our data analysis, we can locate the chiral and spin transition points as $T_{CG}=0.143\pm 0.003$ and $T_{SG} =0.125^{+0.006}_{-0.012}$. We then conclude that the SG transition occurs at a nonzero temperature which is located about 10$\sim $15\%  below the CG transition temperature. Thus, the 3D Heisenberg SG exhibits the spin-chirality decoupling. 

 We also examine the possibility suggested in the previous works that the spin and chiral sectors undergo simultaneously a KT transition \cite{Campos06,LeeYoung07}. From our new data of large sizes $L\leq 32$ covering the temperature range below $T_{g}$, we conclude that such a possibility can now be ruled out. In order to corroborate this conclusion, we also calculate the correlation-length ratio and the Binder ratio for the 2D ferromagnetic {\it XY\/} model, the standard model exhibiting the KT transition, and compare the results with the ones of the 3D Heisenberg SG. Both the correlation-length ratio and the Binder ratio exhibit quite different behaviors between the two models, demonstrating again that the transition of the 3D Heisenberg SG is not of KT-type.

 Recently, Pixley and Young questioned the utility of the Binder ratio in studying the ordering of vector SG models which is characterized by many-component tensorial order parameter \cite{PixleyYoung08}. To examine the validity of this claim, we also simulate the ferromagnetic 3D $O(10)$ Heisenberg model, a model with a large number of order-parameter components $n=10$. By calculating the correlation-length ratio and the Binder ratio of the model, and by comparing them with those of the 3D Heisenberg SG, we conclude that the Binder ratio of the 3D vector SG carries useful information independent of the correlation-length ratio, and should not be regarded as behaving in a trivial manner.

 We analyze the critical properties associated with the CG transition. On the basis of a finite-size scaling analysis taking account of the leading correction-to-scaling, we get estimates of the CG critical exponents as $\nu_{CG}=1.4\pm 0.2$ and $\eta_{CG}=0.6\pm 0.2$, the former being the CG correlation-length exponent and the latter the CG critical-point-decay exponent. In fact, these CG exponents are close to the exponent values reported in earlier works, and also turn out to be very close to the SG exponent values experimentally observed for canonical SGs. In fact, this coincidence gives a strong support to the chirality scenario.

 In order to further probe the nature of the CG ordered state, we also calculate the overlap distribution function and the non-self-averaging parameter (the so-called $A$ parameter) both for the chirality and the spin. The CG ordered state turns out to be non-self-averaging. It also appears to exhibit a peculiar type of replica-symmetry breaking (RSB) which closely resembles the so-called one-step RSB. The behavior of the Binder ratio is fully consistent with such a one-step-like RSB picture. A preliminary account of the simulation was reported in Ref.\cite{VietKawamura}.

 The present paper is organized as follows. In \S 2, we define our model and explain some of the details of our numerical method employed. Particular attention is paid to the issue of thermalization, {\it i.e.\/}, how we check the equilibration which is often crucial in obtaining reliable data. Various physical quantities calculated in our simulations are introduced in \S 3. Then, our MC results are presented in \S 4. Quantities like the specific heat, the local-chirality amplitude, the CG and SG correlation-length ratios, the CG and SG susceptibility, the CG and SG Binder ratios, the CG and SG overlap distribution functions, the CG and SG non-self-averaging parameters {\it etc\/}, are calculated. \S 5 is devoted to a finite-size scaling analysis of the CG critical properties. By analyzing the CG correlation-length ratio and the CG order parameter with taking account of the leading correction-to-scaling, we estimate the CG critical exponents. The character of the CG ordered state is also studied via the Binder ratio, the overlap distribution function and the non-self-averageness parameter. Experimental implications are briefly discussed. Finally, \S 6 is devoted to summary and discussion.

\section{The model and method}
We study an isotropic classical Heisenberg model on a 3D simple-cubic lattice defined by the Hamiltonian
\begin{equation}
{\cal H}=-\sum_{<ij>}J_{ij}\vec{S}_i\cdot \vec{S}_j\ \ ,
\label{eqn:hamil}
\end{equation}
where $\vec{S}_i=(S_i^x,S_i^y,S_i^z)$ is a three-component unit vector at the $i$-th site, and the $<ij>$ sum is taken over all nearest-neighbor pairs. The couplings $J_{ij}$ are random Gaussian variables with zero mean and standard deviation unity. The lattice contains $N=L^{3}$ sites with $L=6, 8, 12, 16, 24, 32$, periodic boundary conditions being applied in all directions.   
\begin{center}
 \begin{table}
  \begin{tabular*} {0.5\textwidth} {@{\extracolsep{\fill}} c c c c c c}
   \hline
   \hline
   $L$ & $N_{s}$ & $N_{T}$ & $N_{MC}$ & $T_{max}$ & $T_{min}$ \\ 
   \hline
   6 & 2000 & 32 & $1\times10^5$ & 0.333 & 0.111 \\
   8 & 2000 & 32 & $1\times10^5$ & 0.333 & 0.111 \\
   12 & 2000 & 32 & $1\times10^5$ & 0.333 & 0.111 \\
   16 & 1500 & 32 & $1\times10^5$ & 0.222 & 0.121 \\
   24 & 1000 & 44 & $1\times10^5$ & 0.222 & 0.133 \\
   32 & 800 & 48 & $3\times10^5$ & 0.209 & 0.133 \\
   \hline
   \hline
  \end{tabular*}
\caption{Various parameters of our Monte Carlo simulations.  $L$ is the system size, $N_{s}$ is the number of samples, $N_{MC}$ is the total number of Monte Carlo steps per spin (our unit Monte Carlo step consists of 1 heat-bath sweep and $L$ over-relaxation sweeps),  $T_{max}$ and $T_{min}$ are the highest and the lowest temperatures used in the temperature-exchange run, and $N_{T}$ is the total number of temperature points. Measurements of physical quantities are made over the latter half of the total $N_{MC}$ Monte Carlo steps, while the former half is discarded for thermalization.}
\label{table1}
\end{table}
\end{center}

 We perform an equilibrium MC simulation by using the single-spin-flip heat-bath method and the over-relaxation method, which are combined with the temperature-exchange technique \cite {Hukushima96}. It has been demonstrated that this method is very effective in reducing the slow dynamics of hard-relaxing systems like SGs \cite{LeeYoung07}. 

The simple cubic lattice consists of two interpenetrating sublattices. We perform the heat-bath sweep sequentially through the sites on one sublattice after another.  After the heat-bath sweep, we repeat the over-relaxation sweeps $M$ times sequentially through the sites on each sublattice \cite {Loison05}. A unit over-relaxation process consists of computing the local field $\vec h_i =\sum_j J_{ij} \vec S_j$ felt by a given spin $\vec S_i$ and reflecting the spin $\vec S_i$ with respect to the local field $\vec h_i$ at this site.
\begin{equation}
\vec {S_i}\rightarrow \vec{S_i}' =-\vec {S_i}+2\frac {(\vec {S_i} \cdot \vec {h_i})}{{h_i^2}}\vec {h_i},
\label{eqn:or}
\end{equation}
where $h_i=|\vec h_i|$.

The combination of one heat-bath sweep and $M$ over-relaxation sweeps constitutes our unit MC step. In our following calculation, the number $M$ is taken as being equal to the system size $L$, {\it i.e.\/}, we take $M=L$. 

 After every MC step, we perform the temperature-exchange trial. The method effectively promotes the system to overcome the free energy barrier characteristic of the spin-glass ordered state. We prepare $N_{T}$ spin configurations with the same interaction coupling, sometimes called ``replicas'', which are located at distinct temperatures distributed in the temperature range between $T_{min}$ and $T_{max}$. The maximum temperature $T_{max}$ needs to be high enough so that the auto-correlation time by the single-spin-flip dynamics is short enough. Then, the temperature-exchange trial is made between the two spin configurations at a pair of neighboring temperatures. 

%
%
In Table. I, we show some of the details of our simulation conditions, including the system size (linear dimension) $L$, the number of independent samples (bond realizations) $N_{s}$ , the number of temperature points used in the temperature-exchange process $N_{T}$, the minimum and maximum temperatures $T_{min}$ and $T_{max}$, and the total number of Monte Carlo steps per spin (MCS) performed per replica. The measurement is made over the last half of the $N_{MC}$ MCS, while the former half is discarded for thermalization. The initial spin configuration is taken to be random.

Error bars are estimated via sample-to-sample fluctuations for linear quantities like the order parameters, and by the jackknife method for non-linear quantities like the Binder ratio and the correlation length ratio.

\begin{figure}[ht]
\begin{center}
\includegraphics[scale=0.9]{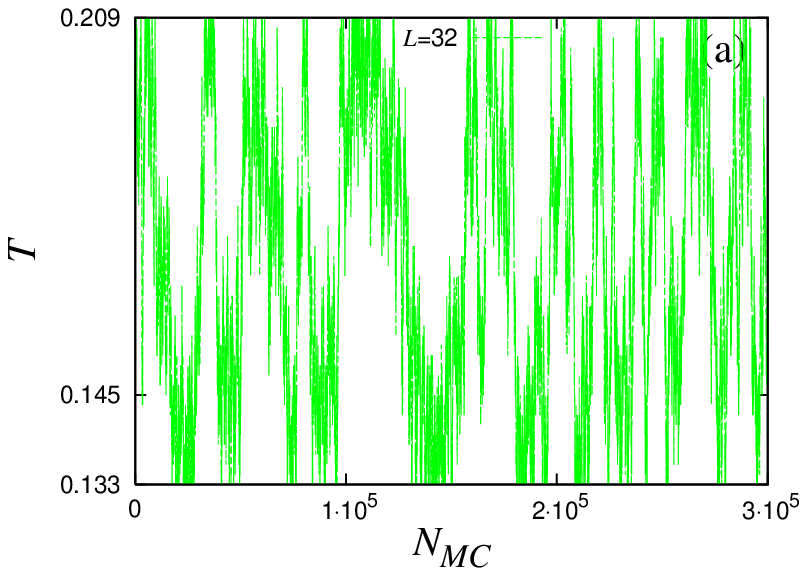}
\includegraphics[scale=0.9]{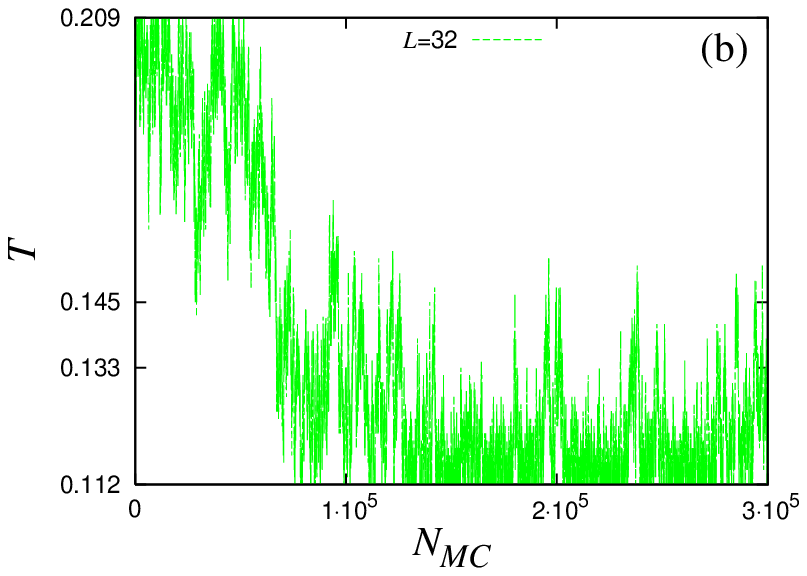}
\end{center}
\caption{
(Color online) An example of typical ``cycling'' behavior of a replica in the temperature-exchange run. The lattice size is $L=32$. In (a), a replica exhibits a frequent cycling between $T_{min}$ and $T_{max}$, where the minimum and maximum temperatures are chosen as $T=T_{min}=0.133$ and  $T=T_{max}=0.209$, which correspond to the values of our final choice for $L=32$.  In (b), a replica exhibits a ``trapping'' behavior, with its move limited in a narrow temperature range over long MC steps. In (b), we set the minimum temperature being lower, $T_{min}=0.112$, while the maximum temperature is the same $T_{max}=0.209$. Note that it often happens that, even when some of replicas exhibit a trapping behavior as shown in (b), other replicas exhibit apparently nice cycling behavior as shown in (a). When even a part of replicas exhibits such a ``trapping'' behavior, the system cannot be regard as being equilibrated. 
}
\end{figure}
\begin{figure}[ht]
\begin{center}
\includegraphics[scale=0.9]{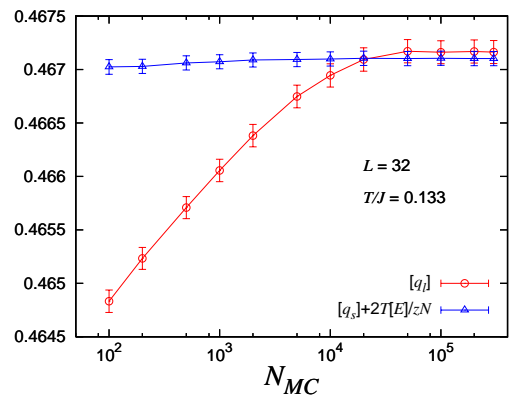}
\end{center}
\caption{
(Color online) Monte Carlo steps dependence of $[q_{l}]$ and $[q_{s}]+\frac {2T} {zN}[E]$  defined by Eqs. (4) and (5). Thermal average is performed over the latter half of the total $N_{MC}$ Monte Carlo steps. In equilibrium, these two quantities should coincide: See Eq. (3). Indeed, the curves of $[q_{l}]$ and of $[q_{s}]+\frac {2T}{zN}[E]$ approach a common value within error bars. The lattice size is $L=32$, and the temperature is $T=T_{min}=0.133$. The sample average is taken for a subset of total samples (150 samples).
}
\end{figure}
\begin{figure}[ht]
\begin{center}
\includegraphics[scale=0.9]{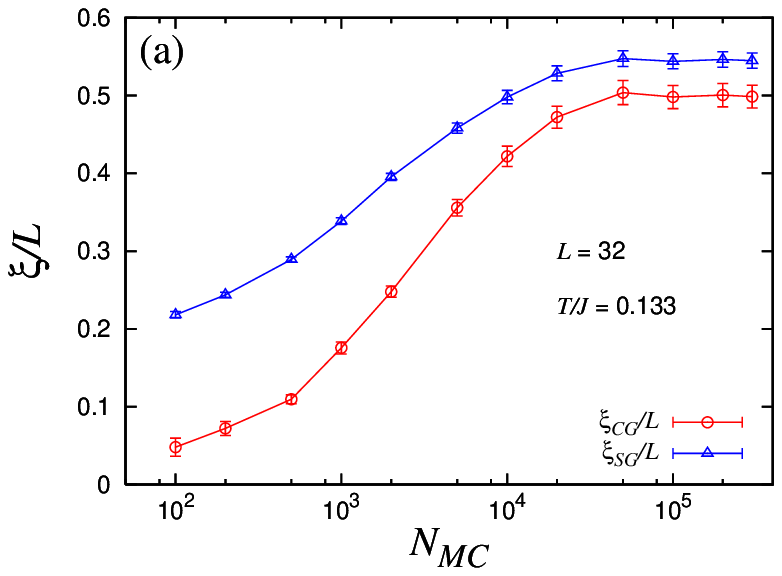}
\includegraphics[scale=0.9]{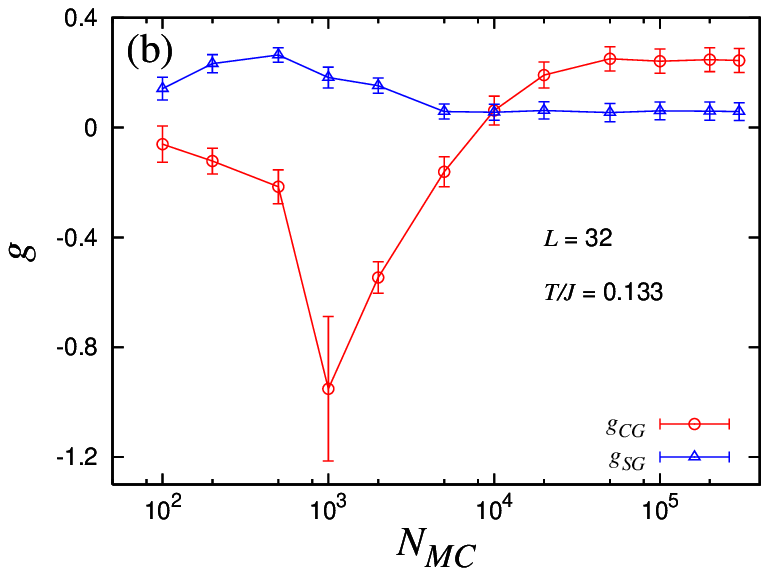}
\includegraphics[scale=0.9]{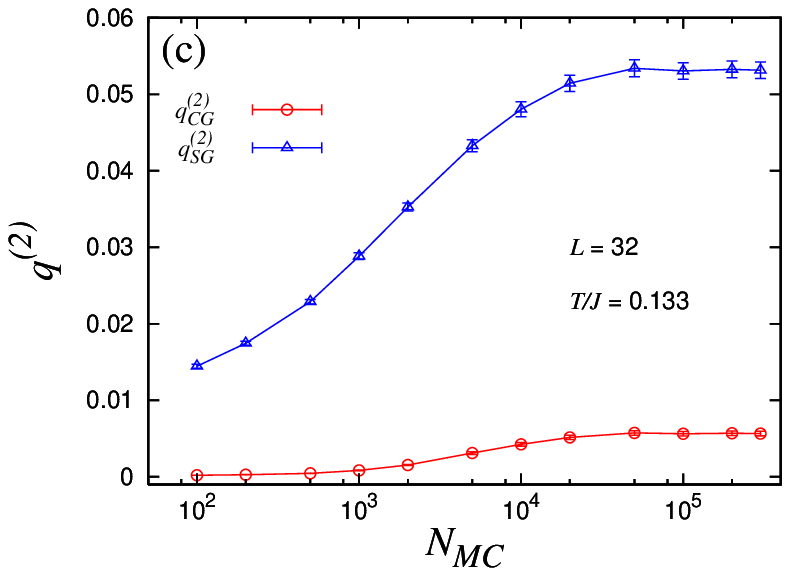}
\end{center}
\caption{
(Color online) The Monte Carlo steps $N_{NC}$ dependence of various physical quantities, $\xi_{CG}/L$ and $\xi_{SG}/L$ (a), $g_{CG}$ and $g_{SG}$ (b), and  $q_{CG}^{(2)}$ and   $q_{SG}^{(2)}$ (c). Thermal average is performed over the latter half of the total $N_{MC}$ Monte Carlo steps. The lattice size is $L=32$, and the temperature is $T=T_{min}=0.133$.  The sample average is taken for a subset of total samples (150 samples).
}
\end{figure}

 One of the most crucial issues in any equilibrium simulation of SGs is to make sure that the system is fully thermalized. In particular, when we use an extended ensemble method like the temperature-exchange method, this point is particularly important, since, if only a part of replicas is not equilibrated, then all others might be affected, and the entire system might not be in equilibrium. Hence, to ensure that the system is fully equilibrated, we need some stringent criteria for equilibration. In the present simulation, we have imposed the following  six conditions for the check of equilibration.

\noindent
1) All of the ``replicas'' move back and forth many times along the temperature axis during the temperature-exchange process (typically more than 10 times) between the maximum and minimum temperature points. A typical cycling pattern of a ``replica'' along the temperature axis during the temperature-exchange process is shown in Fig.1(a) for our largest lattice size $L=32$. We also check that the relaxation due to the single-spin-flip (1 heat-bath sweep plus $M=L$ over-relaxation sweeps) is fast enough at $T=T_{max}$. Both the chiral and spin autocorrelation times at  $T=T_{max}$ turn out to be about 22 MCS for $L=32$, and less than 16 MCS for smaller lattice sizes. This guarantees that different parts of the phase space are sampled in each ``cycle'' of the temperature-exchange process. 

 It sometimes happens for larger lattice size $L$ and the choice of lower $T_{min}$ that the frequent cycling between $T_{min}$ and $T_{max}$ cannot be achieved in a part of replicas: Some of the replicas, often not all, are ``trapped'' in a restricted temperature range in the course of simulation. A typical example of such a ``trapping'' behavior is shown in Fig.1(b) for the case of our largest size $L=32$, where the lowest temperature $T_{min}$ is taken to be $T_{min}=0.112$  considerably lower than our final choice $T_{min}=0.133$. Once such a ``trapping'' occurs for certain replicas, an extremely long time is needed to get out of it, and the system can hardly reach thermal equilibrium. A particularly tricky point here is that, once the trapping occurs for certain replicas, it often takes an extremely long time to get out of it so that various physical quantities appear to converge to ``fake'' stable values. It should also be noted that, even when certain replicas exhibit a trapping behavior as shown in Fig.1(b), other replicas continue to exhibit a nice cycling behavior as shown in Fig.1(a). Yet, if the trapping behavior is observed for a part of replicas, the system cannot be regarded as equilibrated at any temperature between $T_{min}$ and $T_{max}$, since the ergodicity is not satisfied as an extended ensemble. Hence, we pay full attention that such a trapping does not occur in any replica by monitoring all the replicas. Once the trapping as in Fig.1(b) is observed in certain replica at the time scale of, say, $10^5$ MCS, we simply abandon the corresponding temperature set and try the new temperature set {\it for all samples\/}, not just one particular sample where we encountered trapping.

\noindent
2) We check that the relation expected to hold  for the model with Gaussian bond distribution in equilibrium \cite{Katzgraber01}, {\it i.e.\/}, whether the relation
\begin{equation}
{[q_{l}]=[q_{s}]+\frac{2T}{zN}[E]}
\end{equation}
is satisfied in the simulation. Here, $E/N$ is the energy per spin and [$\cdots$] represents an average over the bond disorder. The ``link spin overlap'' $q_{l}$ and the quantity $q_{s}$ are defined by
\begin{equation}
q_{l}=(1/N_{b})\sum_{ij}\langle \vec{S}_i\cdot \vec{S}_j\rangle^{2}, 
\end{equation}
\begin{equation}
q_{s}=(1/N_{b})\sum_{ij}\langle(\vec{S}_i\cdot \vec{S}_j)^{2}\rangle, 
\end{equation}
where $N_{b}=(z/2)$ is the number of nearest-neighbor bonds ($z=6$ the coordination number of the lattice), and $\langle\cdots\rangle$ represents a thermal average. As illustrated in Fig.2 in the case of our largest lattice $L=32$, our data well satisfy Eq. (3) within the error bar. Note that, although this criterion is quite useful, it is only a necessary condition of equilibration, not a sufficient condition, as in case of many other criteria.

\noindent
3) We check that measured physical quantities converge to stable values. As an example, we show in Fig.3 the MC time dependence of several physical quantities, including the CG and SG correlation-length ratios $\xi_{CG}/L$ and $\xi_{SG}/L$,  the CG and SG  Binder ratios $g_{CG}$ and $g_{SG}$, and  the CG and SG glass order parameters $q_{CG}^{(2)}$ and  $q_{SG}^{(2)}$ (to be defined in the next section) for $L=32$ and at $T=T_{min}=0.133$.  All these quantities converge to stable values after some period, indicating that the system has been equilibrated. We note that the relaxation is faster at higher temperatures including the critical regime around $T_{CG}$. When equilibration is not sufficient, the correlation length $\xi$ and the glass order parameter $q^{(2)}$ tend to be smaller than the true values as naturally anticipated, while no such inequality seems to exist for the Binder ratio $g$. As can be seen from Fig.3(b), it sometimes occurs that thermalization of the chirality-related quantity takes more time than that of the spin-related quantity. This might indicate that the chirality is harder to relax than the spin. Hence, one should carefully test the stability of the chirality-related quantities in particular, not only of the spin-related quantities.

\noindent
4) We check that the expected symmetry of the overlap distribution function  $P_{J}(q)$ under the reversal operation $q\rightarrow-q$ holds for each individual sample. Since the global flipping of the spins and of the chiralities is supposed to give a slow mode of the system, this gives a stringent test of equilibration. Again, in our simulations, the symmetry of both spin and chiral $P_{J}(q)$ turns out to be excellent for all individual samples \cite {Marinari01}. Of course, this is again only a necessary condition, not a sufficient condition.

\noindent
5) The equality  between the specific heat computed via the energy fluctuation and the one computed via the temperature difference of the energy, which is expected to hold in any  equilibrium system, is checked.


\noindent
6) We compare our data of the correlation length with the recent data reported by other authors, in the temperature range where common data are available \cite {Campos06,LeeYoung07}.

 We have carefully checked that our data satisfy all the criteria 1)-6) above. 
In this way, we believe that the system has been fully equilibrated in our simulations up to the largest lattice size $L=32$ and down to the lowest temperature $T_{min}$.

\section{Physical quantities}

In this section, we define various physical quantities measured in our simulations and discuss some of their basics and details.

For the Heisenberg spin, the local chirality at the $i$-th site and in the $\mu$-th direction $\chi_{i\mu}$ may be defined for three neighboring Heisenberg spins by a scalar
\begin{equation}
\chi_{i\mu}=
\vec{S}_{i+{\hat{e}}_{\mu}}\cdot
(\vec{S}_i\times\vec{S}_{i-{\hat{e}}_{\mu}}),
\end{equation}
where ${\hat{e}}_{\mu}\ (\mu=x,y,z)$ denotes a unit vector along the $\mu$-th axis. There are in total $3N$ local chiral variables.

First, we define an ``overlap'' for the chirality. In addition to ``replicas'' associated with the temperature-exchange process, we also prepare at each temperature two independent systems 1 and 2 (also called ``replicas'' here) described by the same Hamiltonian (1) with the same interaction set. We simulate these two replicas 1 and 2 in parallel with using different spin initial conditions and different sequences of random numbers.  

The $k$-dependent chiral overlap, $q_\chi(\vec k)$, is defined as an overlap variable between the two replicas 1 and 2 as a scalar
\begin{equation}
q_\chi(\vec k) =
\frac{1}{3N}\sum_{i=1}^N\sum_{\mu=x,y,z}
\chi_{i\mu}^{(1)}\chi_{i\mu}^{(2)}e^{i\vec k\cdot \vec r_i},
\end{equation}
where the upper suffixes (1) and (2) denote the two replicas of the system.

The $k$-dependent spin overlap, $q_{\alpha\beta}(\vec k)$, is defined by a {\it tensor\/} variable between the $\alpha$ and $\beta$ components of the Heisenberg spin,
\begin{equation}
q_{\alpha\beta}(\vec k) = 
\frac{1}{N}\sum_{i=1}^N S_{i\alpha}^{(1)}S_{i\beta}^{(2)}e^{i\vec k\cdot \vec r_i},
\ \ \ (\alpha,\beta=x,y,z).
\end{equation}

In term of the $k$-dependent overlap, the CG and SG order parameters are defined by the second moment of the overlap at a wavevector $k=0$,
\begin{equation}
q_{CG}^{(2)}=\frac {[\langle | q_{\chi}(\vec 0)|^2 \rangle]} {\overline{\chi}^{4}},
\end{equation}
\begin{equation}
q_{SG}^{(2)} = [\langle q_{\rm s}(\vec 0)^2\rangle]\ ,
\ \ \  
q_{\rm s}(\vec k)^2 = \sum_{\alpha,\beta=x,y,z} \left| q_{\alpha\beta}(\vec k) \right| ^2.
\end{equation}
The CG order parameter $q_{CG}^{(2)}$ has been normalized here by the mean-square amplitude of the local chirality,
\begin{equation}
\overline{\chi}^{2}=\frac{1}{3N}\sum_i^N\sum_{\mu}[\langle \chi_{i\mu}^2\rangle],
\end{equation}
which remains nonzero only when the spin has a noncoplanar structure locally. The local-chirality amplitude depends on the temperature and the lattice size only weakly as shown later in Fig.5.

The CG and SG susceptibilities are defined by 
\begin{equation}
\chi_{CG}=3Nq_{CG}^{(2)}\ , \ \ \  \chi_{SG}=Nq_{SG}^{(2)}.
\end{equation}

Finite-size correlation lengths are defined by
\begin{equation}
\xi_L = 
\frac{1}{2\sin(k_\mathrm{m}/2)}
\sqrt{ \frac{ [\langle q(\vec 0)^2 \rangle] }
{[\langle q(\vec{k}_\mathrm{m})^2 \rangle] } -1 },
\end{equation}
for each case of the chirality and the spin, $\xi_{CG}$ and $\xi_{SG}$, where $\vec{k}_{\rm m}=(2\pi/L,0,0)$ with $k_{\textrm{m}}=|\vec k_{\textrm{m}}|$, and the $\mu$-direction in Eq.(6) is taken here being parallel with $\vec k$.

The CG and the SG Binder ratios are defined by
\begin{equation}
g_{CG}=
\frac{1}{2}
\left(3-\frac{[\langle q_{\chi}(\vec 0)^4\rangle]}
{[\langle q_{\chi}(\vec 0)^2\rangle]^2}\right),
\end{equation}
\begin{equation}
g_{SG} = \frac{1}{2}
\left(11 - 9\frac{[\langle q_{\rm s}(\vec 0)^4\rangle]}
{[\langle q_{\rm s}(\vec 0)^2\rangle]^2}\right).
\label{eqn:gs_def}
\end{equation}
These quantities are normalized so that, in the thermodynamic limit, they vanish in the high-temperature phase and gives unity in the non-degenerate ordered state. In the present Gaussian coupling model, the ground state is expected to be non-degenerate so that both $g_{CG}$ and $g_{SG}$ should be unity at $T=0$.

 One can also define the non-self-averageness parameter, or the so-called $A$ parameter \cite {Marinari99}, for the chirality and for the spin by
\begin{equation}
A_{CG} = \frac {[\langle q_{\chi}(\vec 0)^2\rangle ^2]-[\langle q_{\chi}(\vec 0)^2\rangle]^2}{[\langle q_{\chi}(\vec 0)^2\rangle]^2},
\end{equation}
\begin{equation}
A_{SG} = \frac {[\langle q_{\rm s}(\vec 0)^2\rangle ^2]-[\langle q_{\rm s}(\vec 0)^2\rangle]^2}{[\langle q_{\rm s}(\vec 0)^2\rangle]^2}.
\end{equation}

The $A$ parameter becomes nonzero if the CG or SG susceptibility is non-self-averaging. It should be noted here that, even if $q_{SG}^{(2)}$ vanishes (or $\chi_{SG}$ remains finite), $A_{SG}$ could become nonzero if $\chi_{SG}$ is not self-averaging. 

 Sometimes, one also uses the so-called Guerra parameter, or the $G$ parameter \cite {Guerra96}, which is defined  by
\begin{equation}
G_{CG} = \frac {[\langle q_{\chi}(\vec 0)^2\rangle ^2]-[\langle q_{\chi}(\vec 0)^2\rangle]^2}{[\langle q_{\chi}(\vec 0)^4\rangle]-[\langle q_{\chi}(\vec 0)^2\rangle]^2},
\end{equation}
\begin{equation}
G_{SG} = \frac {[\langle q_{\rm s}(\vec 0)^2\rangle ^2]-[\langle q_{\rm s}(\vec 0)^2\rangle]^2}{[\langle q_{\rm s}(\vec 0)^4\rangle]-[\langle q_{\rm s}(\vec 0)^2\rangle]^2}.
\end{equation}

Unlike the $A$ parameter, the $G$ parameter can take a nonzero value even when the ordered state is a trivial one without accompanying an RSB \cite {Bokil99,Marinari98}. In fact, the $G$ parameter is not independent of the Binder ratio $g$ and the $A$ parameter, given by
\begin{equation}
G_{CG} =\frac{1}{2}A_{CG}/(1-g_{CG}),
\end{equation}
\begin{equation}
G_{SG} =\frac {9}{2}A_{SG}/(1-g_{SG}).
\end{equation}
Hence, it should be noticed that, even if there is no SG order in the sense $q_{SG}^{(2)}=0$ and $g_{SG}=0$, $G_{SG}$ could take a nonzero value if $\chi_{SG}$ is non-self-averaging, {\it i.e.\/}, $A_{SG}\neq 0$.


 The chiral-overlap distribution $P(q_{\chi})$ is defined by 
\begin{equation}
P(q_{\chi}^{'})=[\langle \delta(q_{\chi}^{'}-q_{\chi}(\vec 0))\rangle].
\end{equation}
The spin-overlap distribution $P(q_{diag})$ is defined originally in the tensor space with $3\times3=9$ components. To make this quantity more easily visible, one may define the diagonal spin-overlap, which is a trace of the original tensor overlap as \cite {ImaKawa03,HukuKawa05} 
\begin{equation}
P(q_{diag})=[\langle \delta(q_{diag}-\sum_{\mu=x,y,z}q_{\mu \mu}(\vec 0))\rangle].
\end{equation}
In the high-temperature phase, both $P(q_{\chi})$ and  $P(q_{diag})$ should approach the $\delta$-function at $q=0$ in the thermodynamic limit. In the low-temperature phase, $P(q_{\chi})$ should develop two symmetric delta-function peaks at the $q_{\chi}$-values corresponding to the chiral EA order parameter $\pm q_{CG}^{EA}$, while  $P(q_{diag})$ should develop two symmetric delta-function peaks at $1/3$ of the spin EA order parameter $\pm q_{SG}^{EA}$: See Refs.\cite{ImaKawa03} for further details.

\section{ Monte Carlo results}

In this section, we present our Monte Carlo results on the three-dimensional isotropic Heisenberg SG with the random Gaussian coupling.

We first show in Fig.4 the temperature dependence of the specific heat for various lattice sizes $L$. An arrow in the figure indicates the location of the CG transition temperature $T_{CG}$, which will be determined below. As can be seen from the figure, the specific heat depends on the temperature only weakly without any appreciable anomaly.
\begin{figure}[ht]
\begin{center}
\includegraphics[scale=0.9]{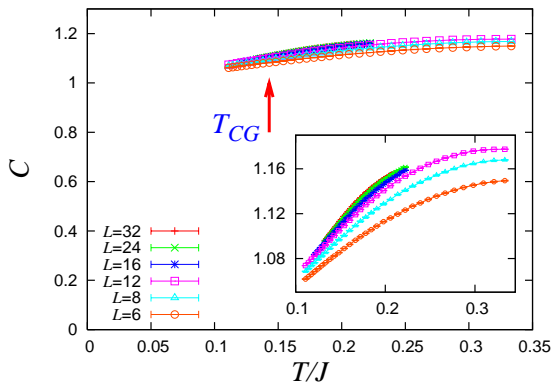}
\end{center}
\caption{
(Color online) The temperature and size dependence of the specific heat per spin. An arrow indicates the location of the chiral-glass transition point. The inset represents a magnified view. In the zero-temperature limit, the specific heat is expected to tend to unity.
}
\end{figure}

 In Fig.5, we show the mean-square local chirality amplitude $\bar \chi^2$ as defined by Eq.(11). Due to the local nature of this quantity, it exhibits only very weak size dependence. It also depends on the temperature weakly, and tends to a nonzero value in the $T\rightarrow 0$ limit, $\bar \chi (T=0)\simeq 0.274$, indicating that the ordered-state spin configuration is locally noncoplanar sustaining a nontrivial scalar chirality.

\begin{figure}[ht]
\begin{center}
\includegraphics[scale=0.9]{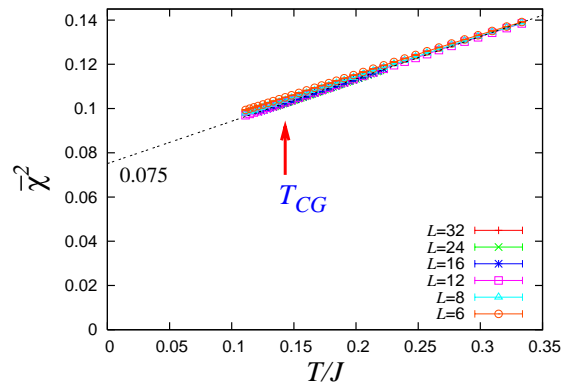}
\end{center}
\caption{
(Color online) The temperature and size dependence of the local-chirality amplitude. An arrow indicates the location of the chiral-glass transition point. In the zero-temperature limit, the local-chirality amplitude is extrapolated to $\bar \chi^2 = 0.075 \pm 0.001$ (or $\bar \chi = 0.274\pm 0.002$): See the broken line in the figure.
}
\end{figure}

 The temperature dependence of the CG and SG order parameters $q_{CG}^{(2)}$ and $q_{SG}^{(2)}$ are shown in Fig.6(a) and Fig.6(b), respectively. The CG order parameter increases more sharply than the SG order parameter, suggesting that the chirality exhibits a stronger ordering tendency than the spin. Meanwhile, more careful analysis of the size dependence is required in determining the transition point, which will be postponed later in this section.
\begin{figure}[ht]
\begin{center}
\includegraphics[scale=0.9]{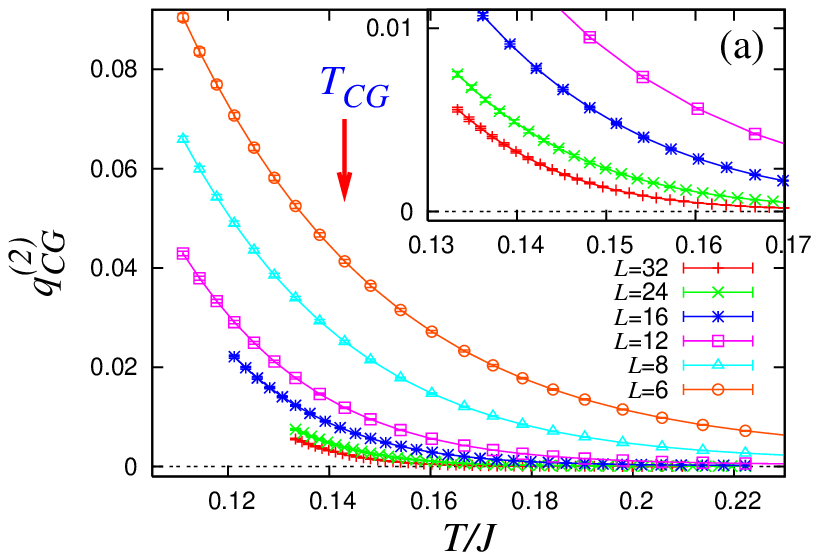}
\includegraphics[scale=0.9]{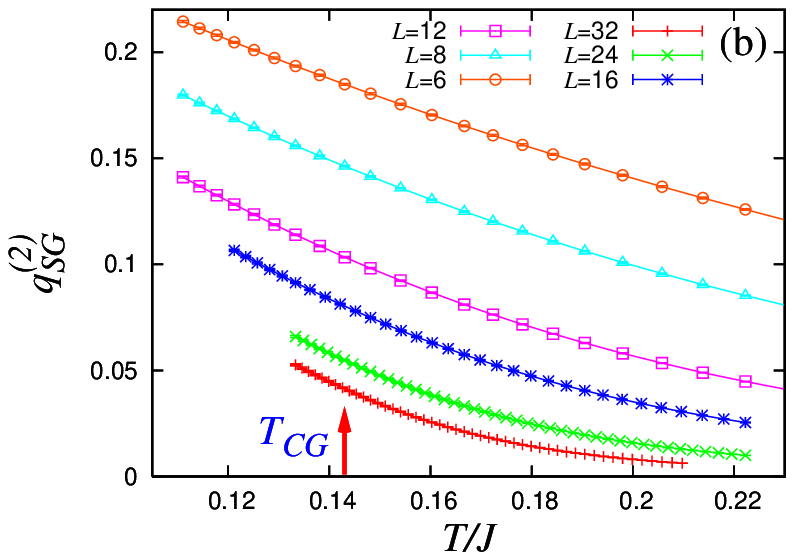}
\end{center}
\caption{
(Color online) The temperature and size dependence of the  chiral-glass order parameter (a), and of the spin-glass order parameter (b).  An arrow indicates the location of the chiral-glass transition point. Inset of Fig.(a) is an enlarged view of the transition region.
}
\end{figure}

The temperature dependence of the CG and SG susceptibilities is shown in Fig.7. In contrast to the SG susceptibility $\chi_{SG}$ which is found to be an increasing function of the lattice size $L$ at all temperature studied, the CG susceptibility $\chi_{CG}$ behaves in this way only at $T/J\lsim0.165$, but exhibits an opposite  size-dependence at $T/J\gsim 0.165$: See the inset of Fig.7(a). Since the chiral susceptibility in the critical regime should be an increasing function of $L$, this observation suggests that the critical region associated with the CG order might be rather narrow. Similar size dependence of $\chi_{SG}$ was also observed in an earlier work \cite{Kawamura95}, and also in the 3D {\it XY\/} SG \cite {Kawamura94}. 
%
\begin{figure}[ht]
\begin{center}
\includegraphics[scale=0.9]{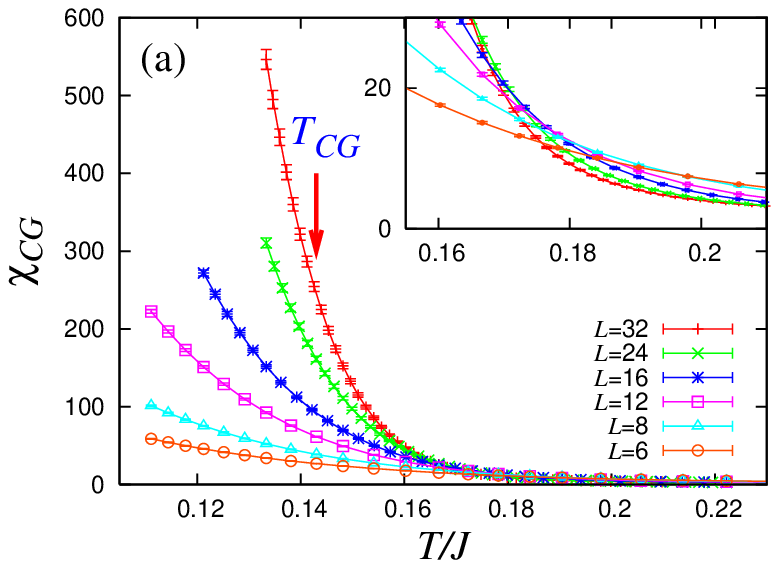}
\includegraphics[scale=0.9]{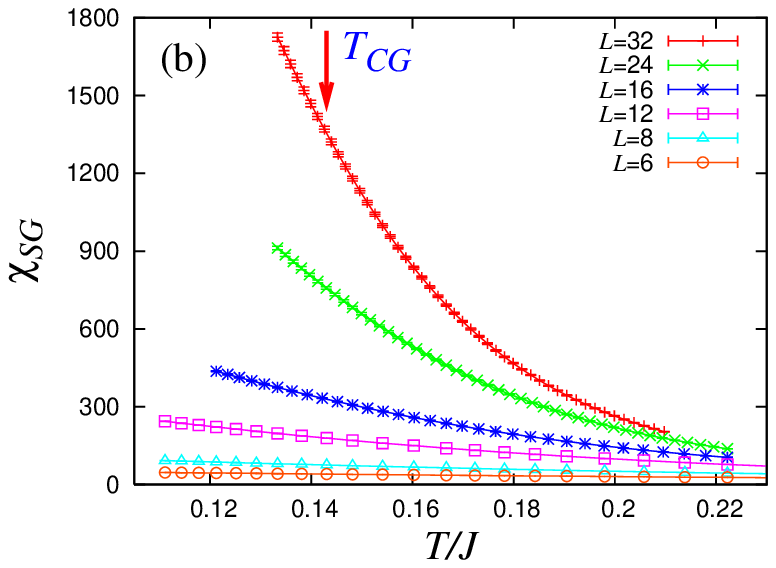}
\end{center}
\caption{
(Color online) The temperature and size dependence of the  chiral-glass susceptibility (a), and of the spin-glass susceptibility (b).  An arrow indicates the location of the chiral-glass transition point.  Inset of Fig.(a) is an enlarged view of the temperature region somewhat higher than the critical regime.
}
\end{figure}

 The temperature dependence of the CG and SG correlation-length ratios, $\xi_{CG}/L$  and $\xi_{SG}/L$, is shown in Figs.8 and 9, an overall behavior in Fig.8 and an enlarged figure in Fig.9. As can be seen from the figures, while the chiral $\xi_{CG}/L$ curves cross at temperatures which are only weakly $L$-dependent, the spin $\xi_{SG}/L$ curves cross at progressively lower temperatures as $L$ increases.

  The present $\xi/L$ data are compared with the data by other authors as follows: Our data for $\xi/L$ are in full agreement with those of Ref.\cite{Campos06} within statistical error bars over the narrow and relatively high-temperature range covered by their data. The data of Ref.\cite{LeeYoung07} for their largest $L$ (on which their claim for ``marginal'' was based) are lower than our present ones and those of Ref.\cite{Campos06} by about 5 to 6 of our $\sigma$ units; this may be a purely statistical effect in view of the limited number of samples measured in Ref.\cite{LeeYoung07}.

\begin{figure}[ht]
\begin{center}
\includegraphics[scale=0.9]{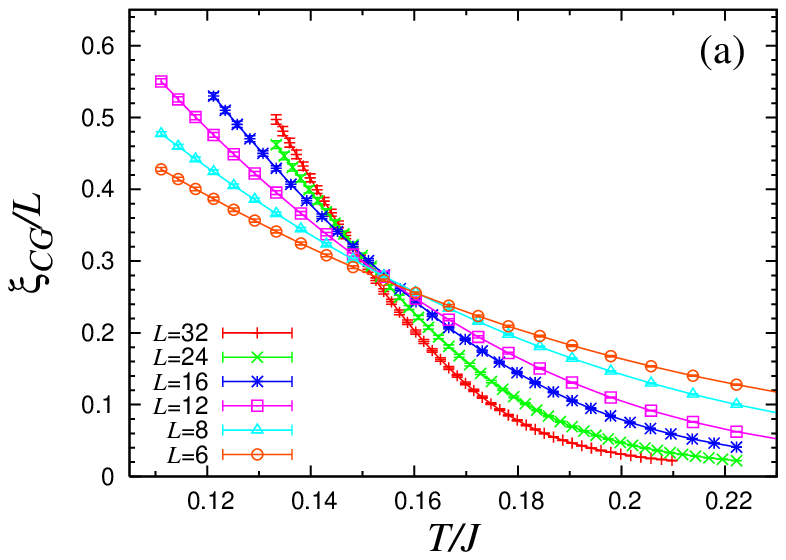}
\includegraphics[scale=0.9]{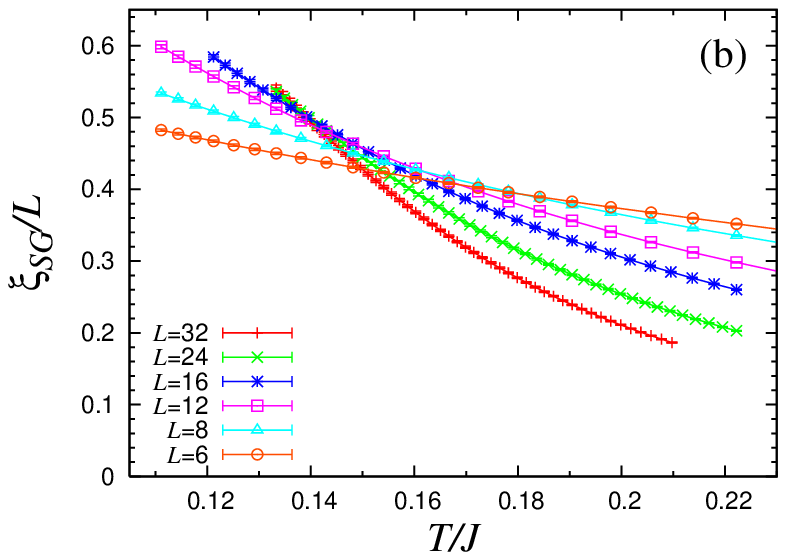}
\end{center}
\caption{
(Color online) The temperature and size dependence of the correlation-length ratio for the chirality (a), and for the spin (b). 
}
\end{figure}
\begin{figure}[ht]
\begin{center}
\includegraphics[scale=0.9]{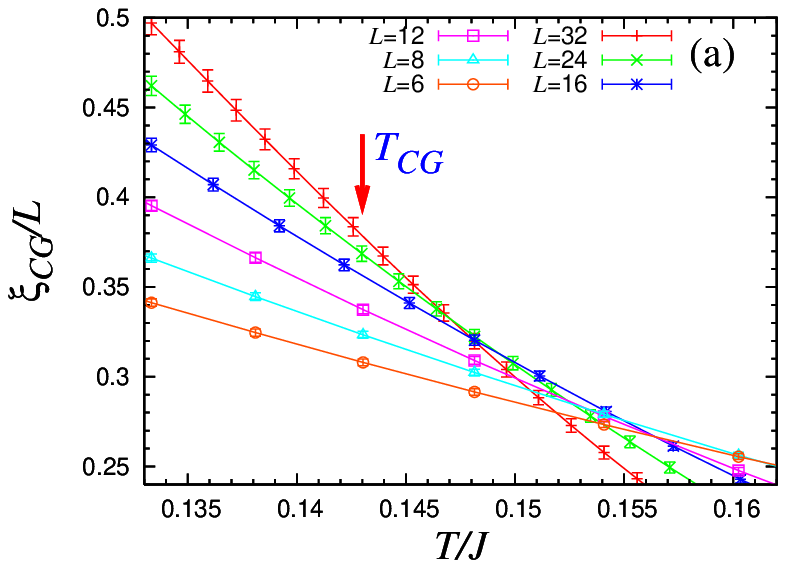}
\includegraphics[scale=0.9]{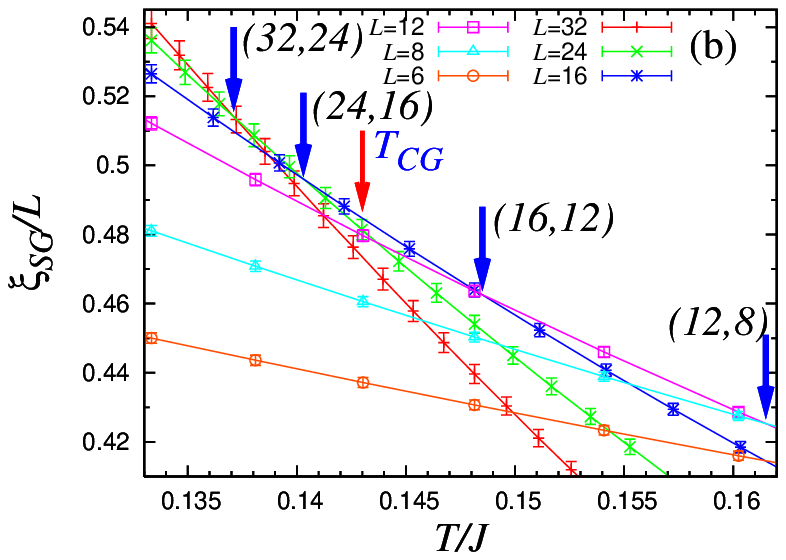}
\end{center}
\caption{
(Color online) Magnified view of the temperature and size dependence of the correlation-length ratio around the transition region for the chirality (a), and for the spin (b).  An arrow indicates the location of the chiral-glass transition point.
}
\end{figure}

 As an other indicator of the transition, we show in Fig.10 the Binder ratios for the chirality (a), and for the spin (b). The chiral Binder ratio $g_{CG}$ exhibits a negative dip which deepens with increasing $L$. The data of different $L$ cross on the {\it negative\/} side of $g_{CG}$. These features indicate a finite-temperature transition in the chiral sector. 

 In order to estimate the bulk CG and SG transition temperatures quantitatively, we plot in Fig.11 the crossing temperatures of $\xi_{CG}/L$ and $\xi_{SG}/L$ and those of $\xi_{SG}$ for pairs of successive $L$ values versus $1/L_{av}$, where $L_{av}$ is a mean of the two sizes. The $L_{av}$-dependence of the dip temperature of $g_{CG}$ is also shown in the figure. Since the data turn out to show an almost linear $1/L_{av}$-dependence, we tried in Ref.\cite{VietKawamura} a simple linear extrapolation of the crossing temperatures $T_{cross}(L)$ and the dip temperature $T_{dip}(L)$ to obtain  $T_{CG}= 0.145\pm 0.005$ and  $T_{SG}= 0.120\pm 0.006$. In the following, we try further elaborate analysis.

\begin{figure}[ht]
\begin{center}
\includegraphics[scale=0.9]{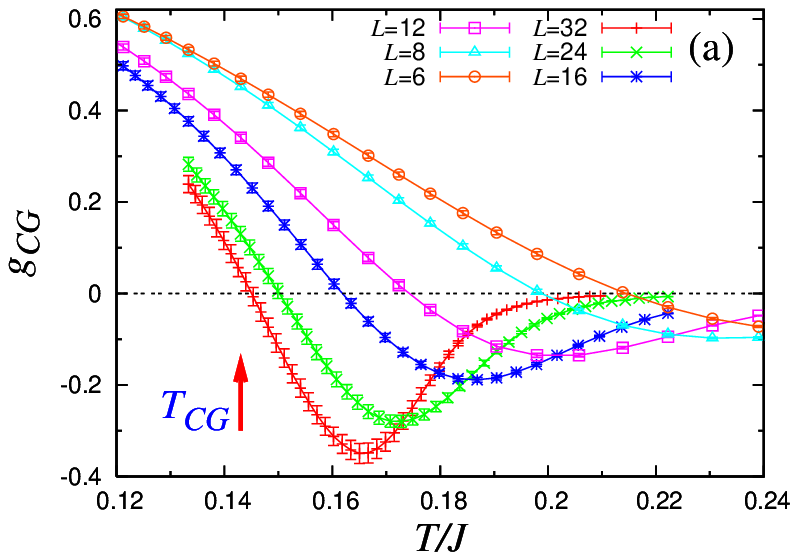}
\includegraphics[scale=0.9]{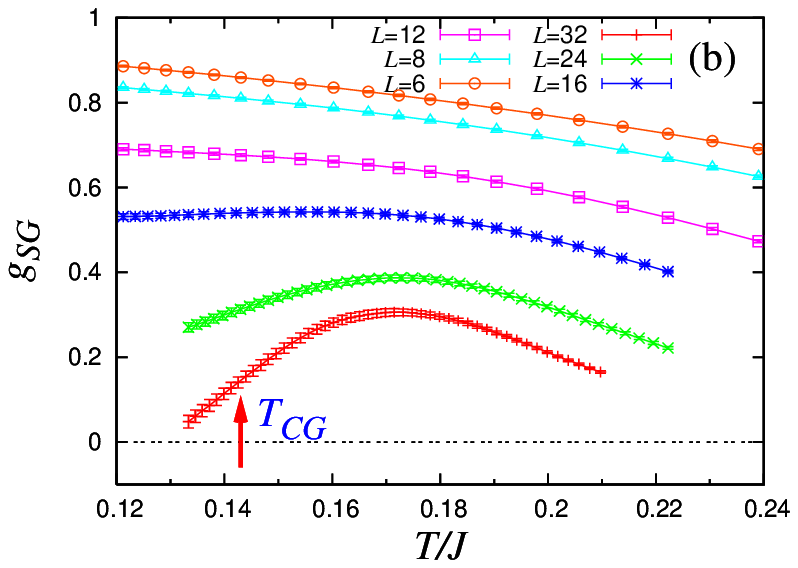}
\end{center}
\caption{
(Color online) The temperature and size dependence of the Binder ratio for the chirality (a), and  for the spin (b). An arrow indicates the location of the chiral-glass transition point.
}
\end{figure}
\begin{figure}[ht]
\begin{center}
\includegraphics[scale=0.9]{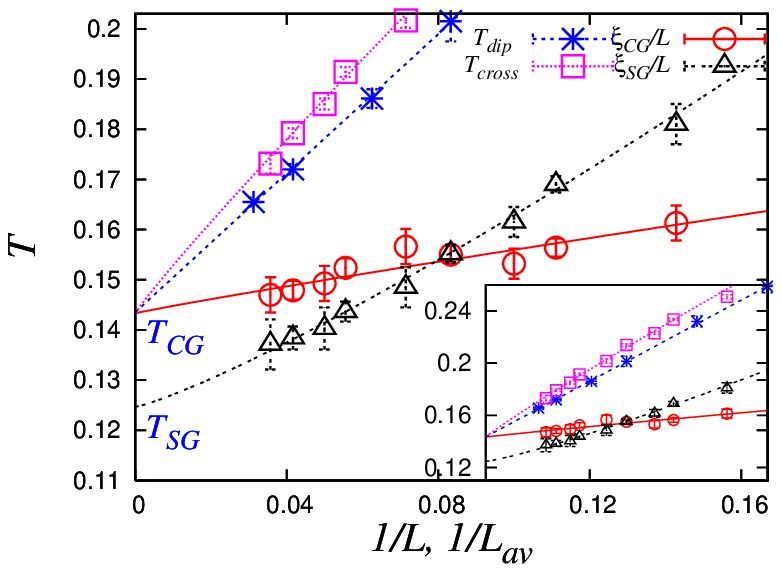}
\end{center}
\caption{
(Color online) The (inverse) size dependence of the crossing temperatures of $\xi_{CG}/L$ and  $\xi_{SG}/L$, the dip temperature $T_{dip}$ and the crossing temperature $T_{cross}$ of $g_{CG}$. Lines represent the fitting curves of the data based on Eq.(24). The spin-glass and chiral-glass transition temperatures are extrapolated to $T_{CG}=0.143\pm 0.003$ and  $T_{SG}=0.125^{+0.006}_{-0.012}$. The inset exhibits a wider range.
}
\end{figure}

 Considerable shift of the crossing temperature with the system size $L$ observed in Fig.11 suggests the relative importance of the correction-to-scaling effect. Generally, one expects 
\begin{equation}
T_{cross}(L)-T_{cross}(\infty)\approx L^{-\theta}, \ \ \  
\theta=\omega+\frac{1}{\nu} , 
\end{equation}
where $\nu$ is the correlation-length exponent and $\omega$ is the leading correction-to-scaling exponent. (Incidentally, Ref.\cite{VietKawamura} quoted $\theta=\omega$, which was inappropriate in the standard notation. This does not affect the subsequent analysis of Ref.\cite{VietKawamura}, though.) Here we perform a joint fit of $T_{cross}(L)$ of both the chiral correlation-length ratio and the chiral Binder ratio, $\xi_{CG}/L$ and $g_{CG}$, to the form Eq.(24), where the CG  transition temperature $T_{CG}=T_{cross}(\infty)$ and the exponent $\theta$ are taken to be common between $\xi_{CG}/L$ and $g_{CG}$. The optimal fit is achieved at $T_{CG}= 0.143$ and  $\theta= 0.93$. To estimate the error bar, we show in Fig.12 the associated $\chi^2$-values of the fit as a function of the assumed $T_{CG}$ and $\theta$ values: In Fig.12(a), the dependence on $T_{CG}$ is shown with optimizing $\theta$ for each $T_{CG}$, while, in Fig.12(b), the dependence on $\theta$ is shown with optimizing $T_{CG}$ for each $\theta$. From these plots, we get estimates  $T_{CG}= 0.143\pm 0.003$ and $\theta= 0.93\pm 0.06$, which turn out to be consistent with our previous estimates  $T_{CG}= 0.145\pm 0.005$ and $\theta\simeq 1$ \cite{VietKawamura}. Our present estimate of $T_{CG}$ also agrees with the one suggested by Campos {\it et al\/}, $T_{CG}\simeq 0.147$ \cite{Campos06}. The fact that we have two independent data sets for $T_{cross}(L)$, one from $\xi_{CG}/L$ and the other from $g_{CG}$, facilitates our estimate of the chiral-glass transition temperature.

\begin{figure}[ht]
\begin{center}
\includegraphics[scale=0.9]{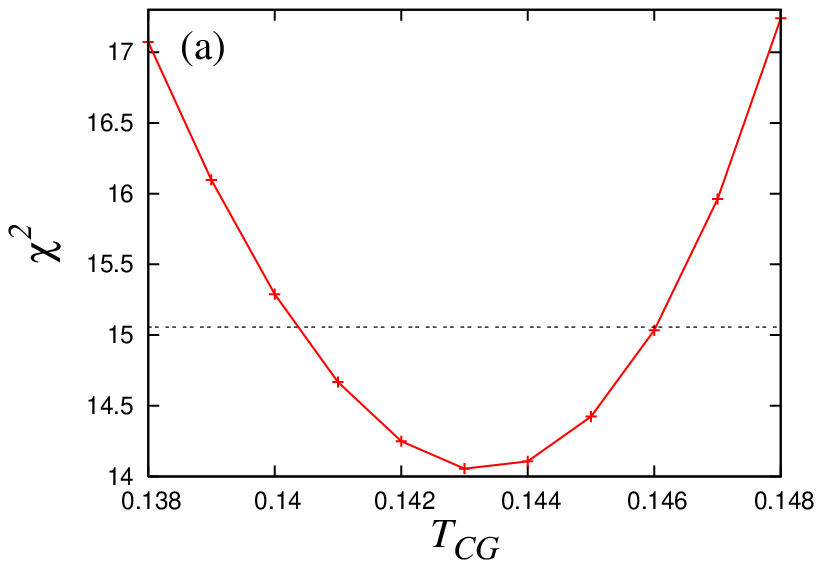}
\includegraphics[scale=0.9]{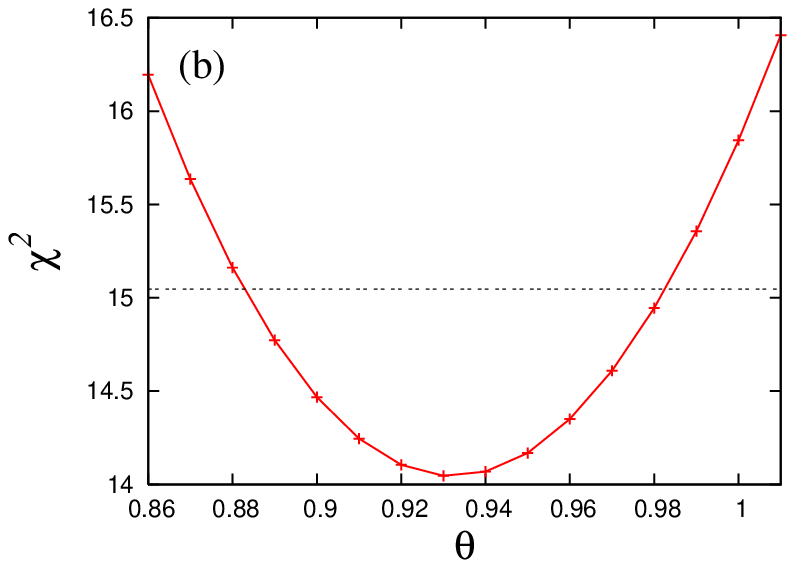}
\end{center}
\caption{
(Color online) The $\chi^2$-value of the fit of $T_{cross}(L)$ to the form Eq.(24) as a function of the assumed $T_{CG}$-value (a), and of the assumed $\theta$-value (b), where other fitting parameters are optimized. The best $\chi^2$-value is obtained at $T_{CG}=0.143$ and $\theta=0.93$. The horizontal straight line corresponds to the $\chi^2$-value equal to the minimum $\chi^2$-value plus unity, which is usually used to estimate error bars.
}
\end{figure}

 We have performed a similar $\chi^2$-analysis based on Eq.(24) also for the crossing temperature of the spin-glass correlation-length ratio $\xi_{SG}/L$. (For the spin, we have only one kind of crossing temperature.) We then get $T_{SG}= 0.125^{+0.006}_{-0.012}$ and $\theta= 1.2\pm 0.35$. The obtained $T_{SG}$ value is consistent within the errors with our previous estimate $T_{SG}= 0.120\pm 0.006$ \cite{VietKawamura}.

 Then, our estimates of $T_{CG}= 0.143\pm 0.003$ and $T_{SG}= 0.125^{+0.006}_{-0.012}$ suggests that $T_{SG}$ is lower than $T_{CG}$ by about 10$\sim $15\%, indicating the occurrence of the spin-chirality decoupling. If we force the chiral crossing-points data to obey $T_{CG}=0.125$, the associated $\chi^2$-value is greater than the optimal value obtained with  $T_{CG}=0.143$ by 25.5, which is significantly greater than the standard error-bar criterion, unity. Likewise, if we force the spin crossing-points data to obey $T_{SG}=0.143$, the associated $\chi^2$-value is greater than the optimal value obtained with  $T_{SG}=0.125$ by 19.3, which is again significantly greater than the standard error-bar criterion, unity. Hence, a simultaneous spin and chiral transition is highly unlikely from our present data.

In Fig.13, we show the ratio of the CG and SG correlation lengths $\xi_{CG}/\xi_{SG}$. The ratio curves of different $L$ intersect. More precisely, for smaller sizes of $L\lsim 12$, this ratio tends to be almost size-independent at lower temperatures indicating that the chiral and spin correlation lengths behave quite similarly \cite{LeeYoung03}. By contrast, for larger sizes of $L\gsim 16$, the ratio curve splays out in the lower temperature regime. This change of behavior of the ratio $\xi_{CG}/\xi_{SG}$ is quite consistent with the  expected size-crossover from the trivial coupling behavior for smaller $L\lsim 12$ to the decoupling behavior for larger $L\gsim 16$ \cite{HukuKawa05}. The crossing point of the ratio curves for our largest $L$ comes around $T\simeq 0.154$, which seems consistent with our estimate above of $T_{CG}\simeq 0.143$. Meanwhile, the ratio itself is still less than unity even at the lowest temperature studied. Still larger lattice size is required to reach the region where this ratio exceeds unity. 

\begin{figure}[ht]
\begin{center}
\includegraphics[scale=0.9]{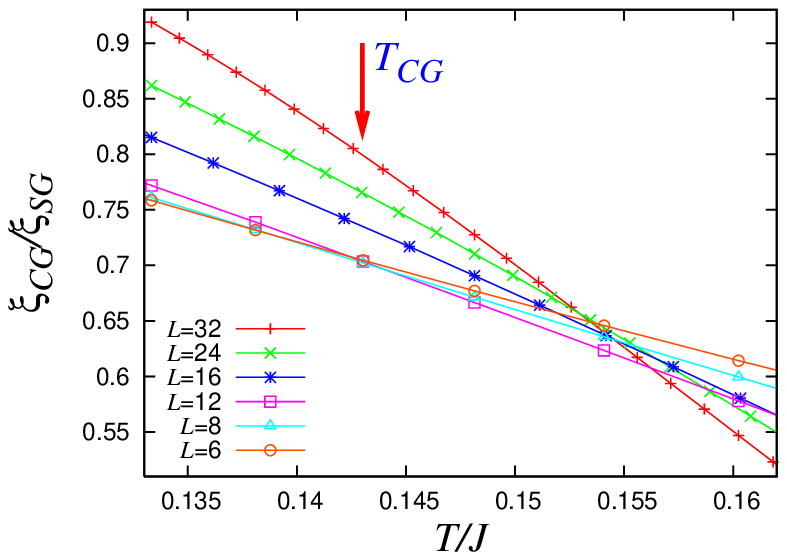}
\end{center}
\caption{
(Color online) The temperature and size dependence of the ratio of the CG correlation-length ratio to the SG correlation-length ratio for various sizes.  An arrow indicates the location of the chiral-glass transition point.
}
\end{figure}

 Recently, Campos {\it et al\/} claimed on the basis of their data of the correlation-length that the chiral and spin sectors undergo simultaneously a KT transition \cite{Campos06}. This interpretation, however, was criticized in Ref.\cite{Campbell07}. More recently, Lee and Young also suggested on the basis of their data of the correlation-length ratios that the system exhibits a ``marginal'' behavior \cite{LeeYoung07}. In view of such recent claims on the model, we further examine here the possibility of a KT-type phase transition. 

 First, we note that, with decreasing the temperature, our data of the CG and SG correlation-length ratios of various $L$ do not merge as expected for the KT transition, but intersect, with crossing points shifting to lower temperature for large $L$. For temperatures below the crossing points, the $\xi_{CG}/L$ and $\xi_{SG}/L$ curves fan out, rather than becoming $L$-independent. In order to make a more direct comparison with the behavior of the KT transition, we calculate the correlation-length ratio $\xi/L$ for the  ferromagnetic 2D {\it XY\/} model, a standard model displaying the KT transition. The result is given in Fig.21 of Appendix A. As can be seen from the figure, $\xi/L$ curves of the 2D {\it XY\/} ferromagnet do not cross at any temperature but merge for larger $L$, becoming asymptotically $L$-independent at temperatures lower than the KT transition temperature. Hence, our present data of either the CG or SG correlation-length ratio shown in Fig.9 are radically different from the one of a typical KT transition shown in Fig.21: Our data of $\xi/L$ curves of the 3D Heisenberg SG are not ``merging''\cite{Campos06} nor ``marginal'' \cite{LeeYoung07}, but splay out.

 In Ref.\cite{Campos06}, Campos {\it et al\/} performed a KT-type scaling with massive logarithmic corrections for the SG correlation-length ratio $\xi_{SG}/L$, and reported that the data exhibited a good scaling. We also tried an exactly same scaling plot with the same logarithmic correction term as performed by Campos {\it et al\/} \cite{Campos06}, and the result is shown in Fig.13. Note that our present data include the low temperature range below $T_{CG}$ which was not covered by the data by Campos {\it et al\/}. (The temperature range where Campos {\it et al\/} reported their scaling plot is indicated by the dashed-line box in our Fig.13.)  As is evident from Fig.13, the KT scaling turns out to be poor, even with invoking a massive logarithmic correction.  We thus conclude that the possibility that the spin and chiral sectors undergo a simultaneous KT-type transition can be ruled out from our present data of the correlation-length ratio.
\begin{figure}[ht]
\begin{center}
\includegraphics[scale=0.9]{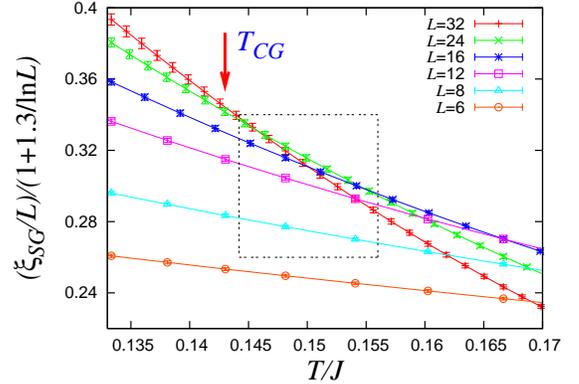}
\end{center}
\caption{
(Color online)  Kosterlitz-Thouless-type scaling plot with a logarithmic correction as performed in Ref.\cite{Campos06} is applied to our data of the spin correlation-length ratio.  An arrow indicates the location of the chiral-glass transition point. The window in the figure exhibits the data range presented in Fig.3b of Ref.\cite{Campos06}. The scaling turns out to be poor.
}
\end{figure}

 The peculiar form of $g_{CG}$ with a negative dip shown in Fig.10(a) is consistent with the occurrence of a one-step-like replica-symmetry breaking (RSB) as suggested by Hukushima and Kawamura \cite{HukuKawa00,HukuKawa05}. This interpretation is corroborated by the form of the   the calculated chiral-overlap distribution below $T_{CG}$ to be shown later in Fig.16(a), which exhibits a prominent central peak at $q_\chi =0$.

 By contrast, the corresponding spin Binder ratio $g_{SG}$ shown in Fig.10(b) does not exhibit a crossing nor a merging in the temperature range studied, suggesting that the SG transition temperature, if any, occurs below $T\simeq 0.13$. Meanwhile, as the size $L$ is increased, $g_{SG}$ develops more and more singular form at low temperature, indicating that the associated overlap distribution significantly changes its shape at low temperature. If one recalls the fact that $g_{SG}$ takes a value unity at $T=0$, $g_{SG}$ is expected to develop a negative dip at a lower $T$ (of $\lsim 0.13$) accompanied by an upturn toward $T=0$. This feature of $g_{SG}$ strongly suggests the occurrence of a SG transition at a nonzero temperature, $T_{SG}\lsim 0.13$.  In order to locate $T_{SG}$ from $g_{SG}$ more directly, however, we need the large-lattice data at lower temperatures.

 In view of the suggestion of the KT-type transition in Refs. \cite{Campos06} and \cite{LeeYoung07}, we further examine the possibility of the KT-type transition via the CG and SG Binder ratios. In order to make a direct comparison with a typical behavior of the KT transition, we calculate the Binder ratio $g$ for the ferromagnetic 2D {\it XY\/} model, and the result is given in Fig.22 of Appendix A. As can be seen from the figure, the $g$ curves of the 2D {\it XY\/} ferromagnet of different $L$ weakly cross at a temperature above the KT transition temperature $T_{KT}$, which gradually tend to $T_{KT}$ in the $L\rightarrow \infty$ limit. Hence, the present data of either the CG or SG Binder ratio shown in Fig.10 are radically different from those of a typical KT transition: Our spin $g_{SG}$ curves do not cross, while our chiral $g_{CG}$ curves exhibit a negative dip. Again, our data of the Binder ratio of the 3D Heisenberg SG are inconsistent with the KT scenario.

 Pixley and Young recently criticized that the Binder ratio might not be an appropriate quantity in studying the ordering of vector SGs, arguing that the large number of order-parameter components ($n=3^2=9$ in the Heisenberg SG) might lead to a trivial Gaussian distribution even below $T_g$ \cite{PixleyYoung08}. To check the validity of such an expectation, we calculate the Binder ratio $g$ of a simple 3D $O(n)$ ferromagnet with large number of $n=10$ components, and the result is given in Fig.23 of Appendix B. As can be seen from the figure, $g$ of 3D $O(10)$ Heisenberg ferromagnet exhibits a clear crossing behavior at the transition temperature $T_c$ and splay out below $T_c$, the behavior characteristic of the standard long-range ordered phase, in spite of the large number of its order-parameter components. Very much similar behavior was also observed in the Binder ratio of the 3D $O(6)$ ferromagnet \cite {Loison99a}. Such a behavior of $g$ is quite different from the one of the 3D Heisenberg SG we observed in Fig.10(b). Hence, the result presents counter-examples to the criticism of Ref.\cite{PixleyYoung08}, demonstrating that the peculiar behavior of $g_{SG}$ observed for the 3D Heisenberg SG in Fig.10(b) should be regarded as a manifestation of essential features of the SG ordering, not mere an artifact due to the large number of order-parameter components. 

 In Fig.15, we show the size dependence of the CG and SG order parameters $q_{CG}^{(2)}$ and $q_{SG}^{(2)}$ on a log-log plot for several temperatures. Straight lines are drawn by fitting the three data points of smaller sizes $L=6$, 8 and 12 at each temperature. As can be seen from Fig.14(a), $q_{CG}^{(2)}$ exhibits an almost linear behavior at a temperature $T=0.148$, an upward curvature characteristic of a long-rage ordered state at lower $T$, and a downward curvature at higher $T$ which should eventually tend to a linear behavior with a slope equal to $-d=-3$ in the disordered phase. Thus, the data of $q_{CG}^{(2)}$ are consistent with our conclusion from the analysis of $\xi_{CG}/L$ and $g_{CG}$ above that the CG transition occurs at $T_{CG}= 0.143\pm 0.003$.

 The SG order parameter $q_{SG}^{(2)}$ exhibits a significantly different behavior, {\it i.e.\/}, it exhibits a downward curvature characteristic of a disordered state at $T=0.148\simeq T_{CG}$, or even at $T=0.133<T_{CG}$. At our lowest temperature $T/J=0.121$ where we could equilibrate only smaller lattices of $L\leq 16$, the data exhibit a near linear behavior up to $L=16$, although it is not clear whether this linear behavior extends to larger $L$. Thus, our data  of $q_{SG}^{(2)}(L)$ are consistent with our conclusion from the analysis of $\xi_{SG}/L$ above that a SG transition occurs at $T_{SG}=0.125^{+0.006}_{-0.012}$, whereas, from the present data of $q_{SG}^{(2)}$ only, we cannot rule out the possibility that $T_{SG}$ is significantly lower than this. Although reliable estimate of the corresponding SG exponents is difficult due the remaining uncertainty in $T_{SG}$, our data of $q_{SG}^{(2)}$ in Fig.14(b) enable us to conclude $\eta_{SG} \lsim -0.30$, which definitely differs from the CG $\eta_{CG}$ value.

\begin{figure}[ht]
\begin{center}
\includegraphics[scale=0.9]{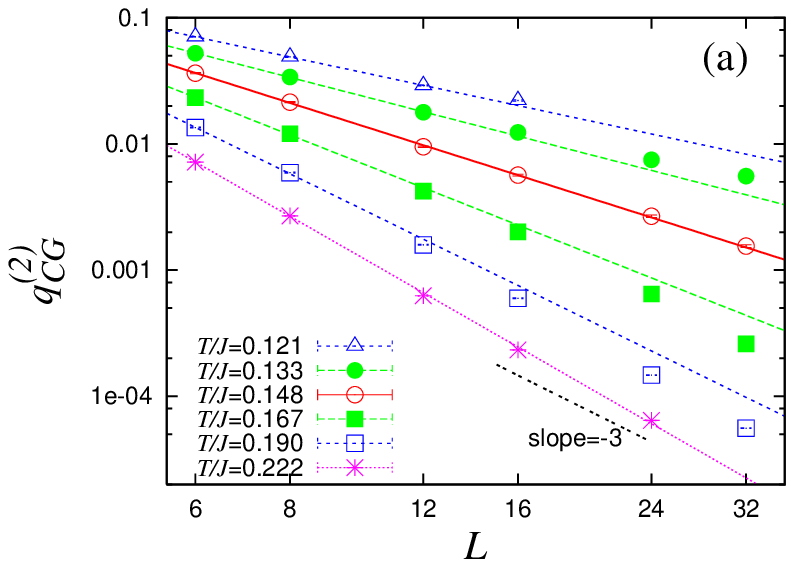}
\includegraphics[scale=0.9]{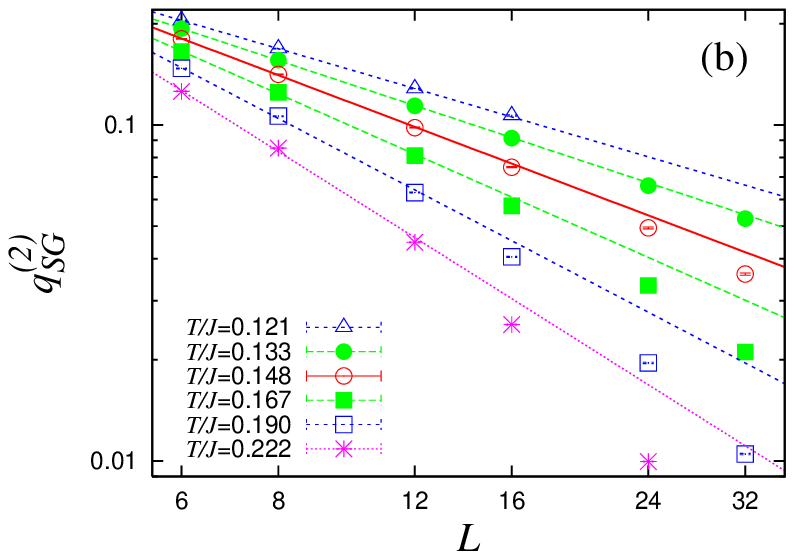}
\end{center}
\caption{
(Color online) The size dependence of the chiral-glass order parameter $q_{CG}^{(2)}$ (a),  and of the spin-glass order parameter $q_{SG}^{(2)}$ (b). Straight lines in the figures are drawn by fitting the three data points of smaller sizes, $L=6,8$ and 12. 
}
\end{figure}
%

%
%

 In Figs.16(a) and (b), we show the chiral-overlap distribution function (a) and the diagonal-spin-overlap distribution function (b) at a temperature $T=0.133$ which lies below $T_{CG}$  but above $T_{SG}$. As one can see from Fig.16(a), the chiral-overlap distribution function $P(q_{\chi})$ displays a central peak at $q_{\chi}=0$ for $L\geq12$ which grows with increasing $L$. In addition to the central peak, there exist small side peaks located at the $q$-values corresponding to the CG EA order parameter $\pm q_{CG}^{EA}$, though these side peaks are weak and look like ``shoulders'' at this temperature. At temperatures higher than $T_{CG}$, $P(q_\chi)$ exhibits only a single Gaussian peak at $q_\chi =0$ without ``shoulders'' even for smaller lattices. The behavior of $P(q_{\chi})$ observed here is similar to the one reported before for the 3D Heisenberg SG with the Gaussian coupling \cite {HukuKawa00}, the 3D Heisenberg SG with the binary coupling \cite {HukuKawa05} and related Heisenberg SG models \cite {KawaIma01,ImaKawa03}. We note that the side peaks of $P(q_\chi)$ were more clearly visible in Refs.\cite{HukuKawa00} and \cite{HukuKawa05}. The form of the overlap distribution characterized by a central peak coexisting with side peaks is the one common to systems exhibiting the so-called one-step RSB. The observed feature of $P(q_{\chi})$ is also consistent with the existence of a negative dip in the CG Binder parameter $g_{CG}$ and with the crossing of $g_{CG}$ occurring on the negative side as discussed before. We note that a one-step feature was also suggested from the study of the fluctuation-dissipation ratio of the 3D Heisenberg SG based on off-equilibrium simulations \cite{Kawamura03}.

 Now we turn to the diagonal-spin-overlap distribution function $P(q_{diag})$ shown in Fig.16(b). Although  $P(q_{diag})$ exhibits a faint double-peak structure or the near flat-peak structure for smaller sizes of $L\leq 12$,  it exhibits  for larger sizes of $L\geq16$ only a single peak located at $q_{diag}=0$, which grows with increasing $L$, without any other appreciable peak structure. This is in contrast to the triple-peak structure observed in the chiral-overlap distribution function $P(q_{\chi})$ of Fig.16(a), which is peaked at $q_{\chi}=0$ and  $\pm q_{CG}^{EA}$. It is also in contrast to the double-peak structure observed in the spin-overlap distribution describing the ordered state of the mean-field Heisenberg SK model, which is peaked at $q=\pm\frac {1}{3}q^{EA}$ \cite{ImaKawa03}. The absence of any divergent peak at nonzero $q_{diag}$ for larger $L$ suggests that the model is in a SG disordered state, at least at a temperature $T=0.133$. This conclusion is consistent with our previous conclusion from the spin correlation-length ratio $\xi_{SG}/L$, the spin Binder ratio $g_{SG}$ and the spin-glass order parameter $q_{SG}^{(2)}$. The appearance of a faint double-peak structure or a near flat-peak structure for smaller sizes of $L\leq 12$ might be interpreted as a size-crossover from the small size SG pseudo-order to the large size SG disorder resulting from the expected spin-chirality coupling-decoupling behavior.

\begin{figure}[ht]
\begin{center}
\includegraphics[scale=0.9]{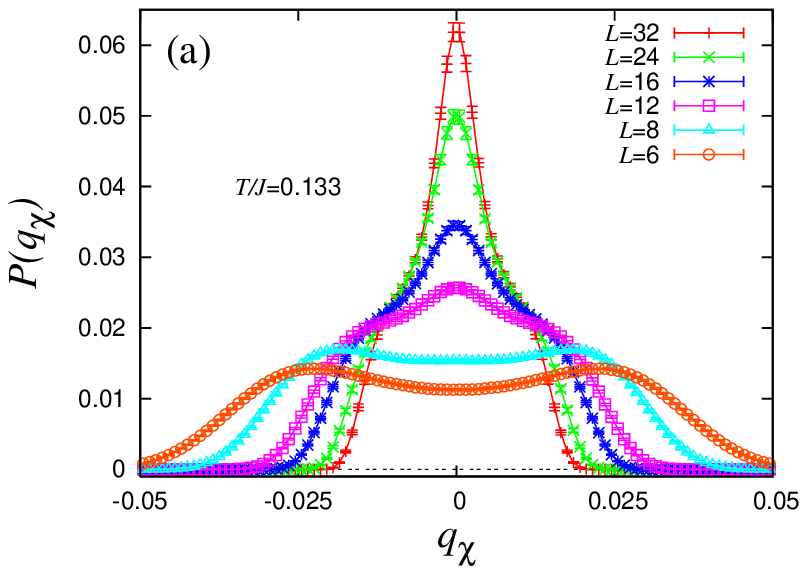}
\includegraphics[scale=0.9]{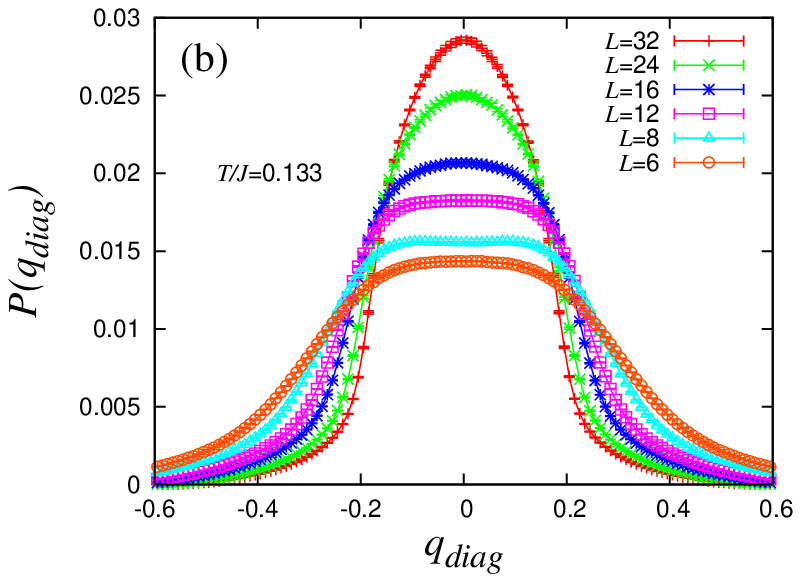}
\end{center}
\caption{
(Color online) The overlap distribution function for the chirality (a), and for the spin (b), at a temperature $T=0.133$.
}
\end{figure}

 In Fig.17, we show the temperature dependence of the non-self-averageness $A$ parameters for the chirality (a) and for the spin (b), respectively.  Although the data are rather noisy due to large sample-to-sample fluctuations, the chiral $A_{CG}$ parameter of different $L$ show a crossing and a prominent peak around the expected $T_{CG}$. This behavior of $A_{CG}$ resembles the one observed in the 3D Ising SG with the Gaussian coupling \cite{Palassini03}, the one of the 3D Heisenberg SG with the binary coupling \cite {HukuKawa05}, and the one of certain mean-field SG models \cite {HukuKawa000,Picco01}. At high temperatures, both $A_{CG}$ and $A_{SG}$ tend to zero with increasing $L$, demonstrating that the system is self-averaging in this regime. Near $T=T_{CG}$, the peak hight of $A_{CG}$ increases with increasing $L$, indicating that the system is non-self-averaging at $T_{CG}$, whereas below $T_{CG}$, the $A_{CG}$ still stays at nonzero value with increasing $L$, indicating that the CG ordered state is non-self-averaging. These findings, combined with the peculiar shape of $P(q_{\chi})$, suggest that the CG ordered phase accompanies an RSB with a non-self-averaging character. 

 By contrast, the spin $A_{SG}$ parameter does not exhibit a peak at any temperature, but exhibits a crossing which occurs slightly above $T_{CG}$ for the range of sizes studied here. Below the crossing temperature, $A_{SG}$ tends to increase with $L$, suggesting that the $\chi_{SG}$ becomes non-self-averaging. Although the crossing of $A_{SG}$ is certainly a signature of a phase transition, it does not necessarily mean the occurrence of the standard SG transition characterized by a nonzero $q_{SG}^{(2)}$ or by the divergence of $\chi_{SG}$. As mentioned in \S 3, a nonzero $A_{SG}$ persisting in the $L\rightarrow \infty $ limit simply means that the SG susceptibility $\chi_{SG}$ is non-self-averaging. Below the CG transition temperature, one expects that the SG order parameter is still Gaussian distributed around zero with the width corresponding to a finite SG susceptibility $\chi_{SG}$, while the width $\chi_{SG}/\sqrt{N}$ exhibits sample-to-sample fluctuations leading to the non-self-averaging (but finite) $\chi_{SG}$. The latter is a natural consequence of the phase-space narrowing which should inevitably occur in the CG state exhibiting a one-step-like RSB. Of course, in the thermodynamic limit, the width $\chi_{SG}/\sqrt{N}$ vanishes yielding a $\delta$-function located at $q_{\mu\nu} =0$ characteristic of the spin disordered state. Hence, the crossing of $A_{SG}$ curves and a nonzero-value of $A_{SG}$ remaining below $T_{CG}$ are fully compatible with the absence of the standard SG long-range order below $T_{CG}$, which is consistent with our present observation of $T_{CG}>T_{SG}$.
\begin{figure}[ht]
\begin{center}
\includegraphics[scale=0.9]{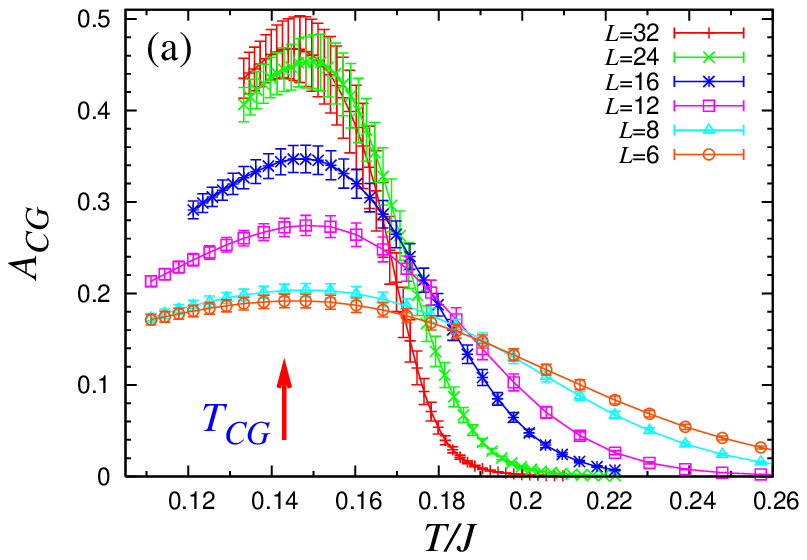}
\includegraphics[scale=0.9]{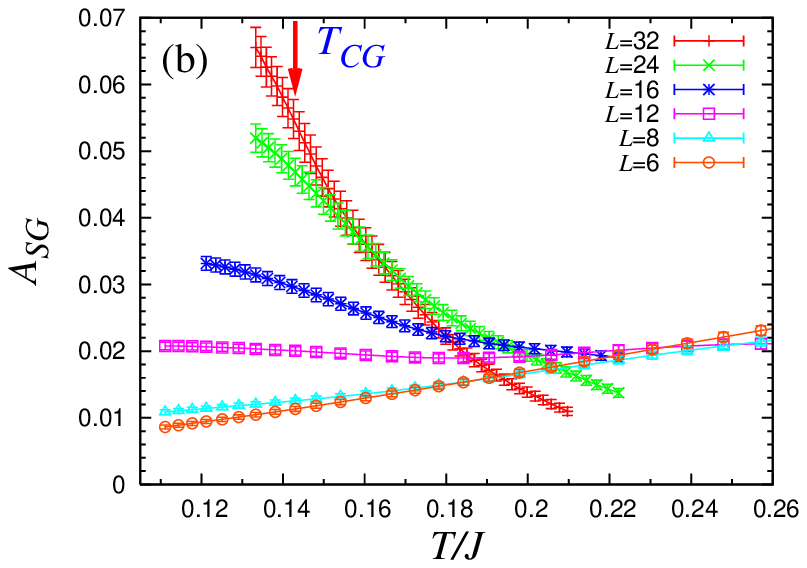}
\end{center}
\caption{
(Color online) The temperature and size dependence of the non-self-averageness $A$ parameter for the chirality (a), and for the spin (b). An arrow indicates the location of the chiral-glass transition point.
}
\end{figure}

 Fig.18 exhibits the temperature dependence of the $G$ parameters for the chirality (a), and for the spin (b). As can be seen from these figures, both the CG and SG $G$ parameters exhibit a crossing near $T_{CG}$, suggestive of a phase transition. As mentioned in \S 3, the $G$ parameter can be written by the $A$ parameter and the Binder ratio $g$ as in Eqs. (20) and (21).  As mentioned, a nonzero $G_{SG}$ occurs once $\chi_{SG}$ becomes non-self-averaging, {\it i.e.\/}, $G_{SG}\neq 0$, even if there is no SG long-range order, {\it i.e.\/}, $q_{SG}^{(2)}=0$ and $g_{SG}=0$: See Eq.(21). Hence, the occurrence of a crossing in $G_{SG}$ at $T_{CG}$ is entirely consistent with our observation of $T_{SG}<T_{CG}$. We also note that the data of $G$ are rather noisy with large error bars as compared with certain other quantities like the correlation-length ratio and the Binder ratio. Therefore, this quantity may not be well suited to an accurate estimate of the transition temperature $T_{g}$. The same suggestion was also made by Ballesteros {\it et al\/}  \cite {Ballesteros00} and by Palassini {\it et al\/} \cite{Palassini03} for the case of the $3D$ Ising SG.  
\begin{figure}[ht]
\begin{center}
\includegraphics[scale=0.9]{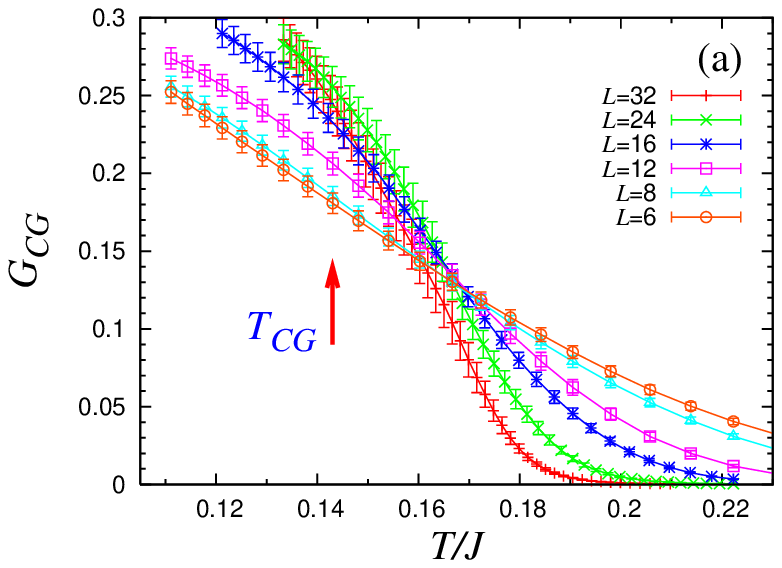}
\includegraphics[scale=0.9]{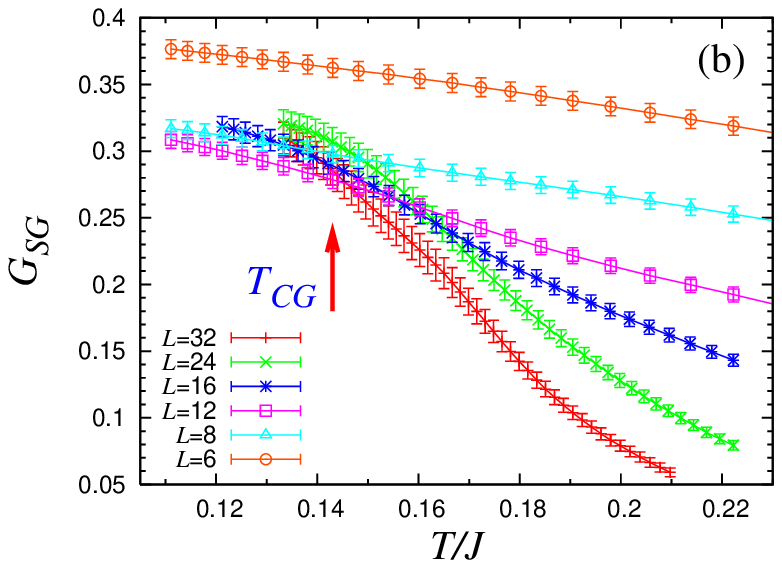}
\end{center}
\caption{
(Color online) The temperature and size dependence of the $G$ parameter for the  chirality (a), and for the spin (b).  An arrow indicates the location of the chiral-glass transition point.
}
\end{figure}

\section{Critical properties of the chiral-glass transition}

 In this section, we study the critical properties of the CG transition on the basis of a finite-size scaling analysis of our data of the CG susceptibility and the CG correlation-length ratio. 

 From our analysis in the previous section, we fix the CG transition temperature to $T_{CG}=0.143$ in this section. Our analysis in \S 4 already suggested the presence of a significant correction-to-scaling term. Hence, we will try  in this section to examine  the effect of the correction-to-scaling, by setting the leading correction-to-scaling exponent to $\theta=\omega+\frac{1}{\nu}=0.93$ as was evaluated in \S 4. We then estimate the two independent critical exponents characterizing the CG transition, {\it i.e.\/}, the CG correlation-length exponent $\nu_{CG}$ and the CG critical-point-decay exponent $\eta_{CG}$.

 The standard finite-size scaling forms for the correlation-length ratio $\xi_{CG}/L$ and for the CG susceptibility ${\chi}_{CG}$ are given by,
\begin{equation}
\frac {\xi_{CG}}{L}=\tilde  X ((T-T_{CG})L^{1/\nu_{CG}}),
\end{equation}
\begin{equation}
\chi_{CG}=L^{2-\eta_{CG}}\tilde Y ((T-T_{CG})L^{1/\nu_{CG}}),
\end{equation}
where $\tilde X$ and $\tilde Y$ are appropriate scaling functions.

For the CG susceptibility $\chi_{CG}$, we obtain a reasonably good scaling by the two-parameter fits with $\nu_{CG}=1.3\pm 0.2$ and $\eta_{CG}=0.7\pm 0.2$. The error bar quoted here and below is estimated by examining by eyes the quality of the fit with varying the fitting parameters.    

For the CG correlation-length ratio $\xi_{CG}/L$, our data shown in Fig.9 have no common crossing point, indicating that the correction-to-scaling term is playing a significant role. Hence, we perform the scaling analysis of $\xi_{CG}/L$ with including the correction-to-scaling term,

\begin{equation}
\frac {\xi_{CG}}{L}=\tilde X ((T-T_{CG})L^{1/\nu_{CG}})(1+aL^{-\omega}),
\end{equation}
where $a$ is a numerical constant, and $T_{CG}$ and $\omega$ are set $T_{CG}=0.143$ and $\omega+\frac{1}{\nu}=0.93$ as mentioned above. The resulting best scaling plot is shown in Fig.19(a), to yield $\nu_{CG}=1.4\pm 0.2$. Note that by including the correction-to-scaling term we can obtain quite a good scaling, which has never been achieved unless we include the correction-to-scaling term in the analysis. This value of $\nu_{CG}=1.4$ happens to be close to the one obtained from $\chi_{CG}$ without invoking the correction-to-scaling term.  

 We also try a similar scaling analysis for $\chi_{CG}$ taking account of the correction-to-scaling term based on the form,
\begin{equation}
\chi_{CG}=L^{2-\eta_{CG}}\tilde Y((T-T_{CG})L^{1/\nu_{CG}})(1+aL^{-\omega}).
\end{equation}
The resulting best scaling plot is given in Fig.19(b). The exponent estimates are $\nu_{CG}=1.4\pm 0.2$ and $\eta_{CG}=0.6\pm 0.2$.

 Recently, Campbell {\it et al\/} proposed an extended version of the standard finite-size scaling method, which might allow one to extend the scaling regime to a wider temperature range \cite {Campbell06}. In this method, one takes an appropriate matching between the data in the critical regime and those in the higher temperature regime to extend the scaling regime. Campbell {\it et al\/} demonstrated that the method worked well for the 3D Ising SG \cite {Campbell06}. The relevant scaling forms are given by
\begin{equation}
\frac {\xi_{CG}}{L}=\tilde  X ((1-\frac {T_{CG}^{2}}{T^{2}})(LT)^{1/\nu_{CG}}),
\end{equation}
\begin{equation}
\chi_{CG} = (LT)^{2-\eta_{CG}} \tilde Y ((1-\frac {T_{CG}^{2}}{T^{2}})(LT)^{1/\nu_{CG}}),
\end{equation}
for $\xi_{CG}/L$ and $\chi_{CG}$, respectively.

 Since this extended scaling method does not take care of the singular correction-to-scaling, it does not serve in itself to improve the quality of the scaling plot of $\xi_{CG}/L$ unless the singular correction-to-scaling term is invoked. If we apply the extended scaling form to $\chi_{CG}$, we get a reasonably good scaling with $\nu_{CG}=1.5 \pm 0.2$ and $\eta_{CG}=0.7\pm 0.2$, which are close to the values obtained based on the standard scaling form Eq.(26). 

\begin{figure}[ht]
\begin{center}
\includegraphics[scale=0.9]{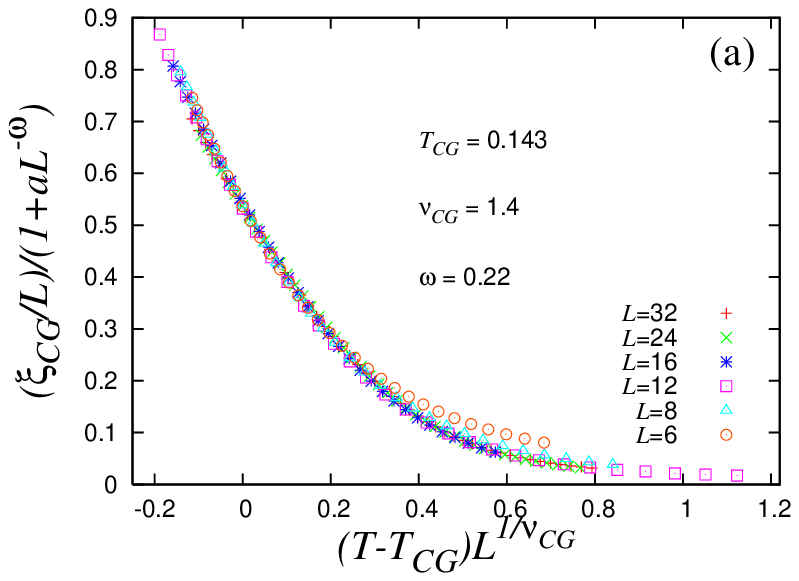}
\includegraphics[scale=0.9]{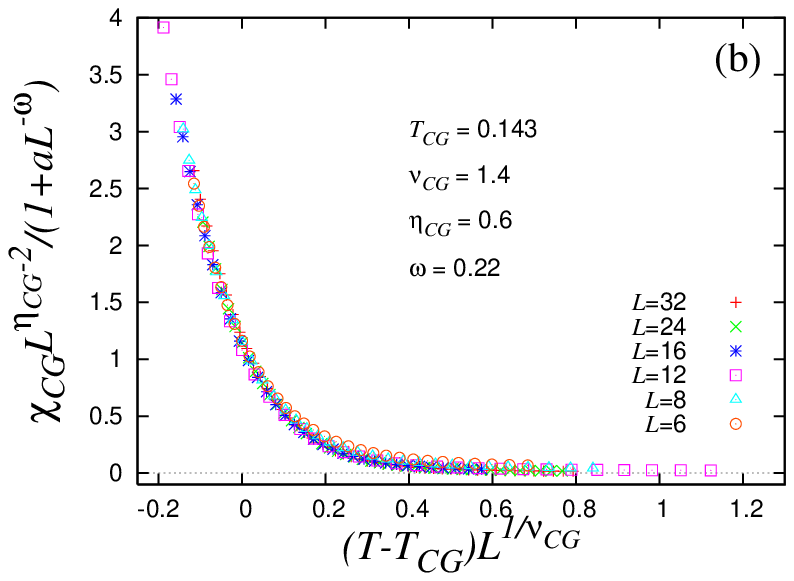}
\end{center}
\caption{
(Color online) Standard finite-size-scaling plots of the chiral-glass correlation-length ratio $\xi_{CG}/L$ (a), and of the chiral-glass susceptibility $\chi_{CG}$ (b), where the correction-to-scaling effect is taken into account. The chiral-glass transition temperature and the leading correction-to-scaling exponents are fixed to $T_{CG}=0.143$ and $\omega+\frac{1}{\nu}=0.93$ as determined in \S 4. The best fit for $\xi_{CG}/L$ is obtained with $\nu_{CG}=1.4$, while that for $\chi_{CG}$ is obtained with $\nu_{CG}=1.4$ and $\eta_{CG}=0.6$.
}
\end{figure}

 It is also possible to apply this extended finite-size scaling both to $\xi_{CG}/L$ and $\chi_{CG}$ with including the correction-to-scaling term. The appropriate scaling forms are given by
\begin{equation}
\frac {\xi_{CG}}{L}=\tilde X ((1-\frac {T_{CG}^{2}}{T^{2}})(LT)^{1/\nu_{CG}})(1+aL^{-\omega}),
\end{equation}
\begin{equation}
\chi_{CG}= (LT)^{2-\eta_{CG}} \tilde  Y ((1-\frac {T_{CG}^{2}}{T^{2}})(LT)^{1/\nu_{CG}})(1+aL^{-\omega}).
\end{equation}
The resulting best scaling plots are given in Fig.20(a) for $\xi_{CG}/L$, and in Fig.20(b) for $\chi_{CG}$. As can be seen from the figures, the quality of the fit is quite good, slightly better  in a wider temperature region than the one obtained from the standard finite-size scaling. The exponent estimates are  $\nu_{CG}=1.5\pm 0.2$ from $\xi_{CG}/L$, and $\nu_{CG}=1.5\pm 0.2$ and $\eta_{CG}=0.6\pm 0.2$ from $\chi_{CG}$.  The value of $\nu_{CG}=1.5$ is slightly higher than the corresponding value  $\nu_{CG}=1.4\pm 0.2$ obtained from the standard finite-size scaling of $\xi_{CG}/L$, but they are fully compatible within the error bar.

\begin{figure}[ht]
\begin{center}
\includegraphics[scale=0.9]{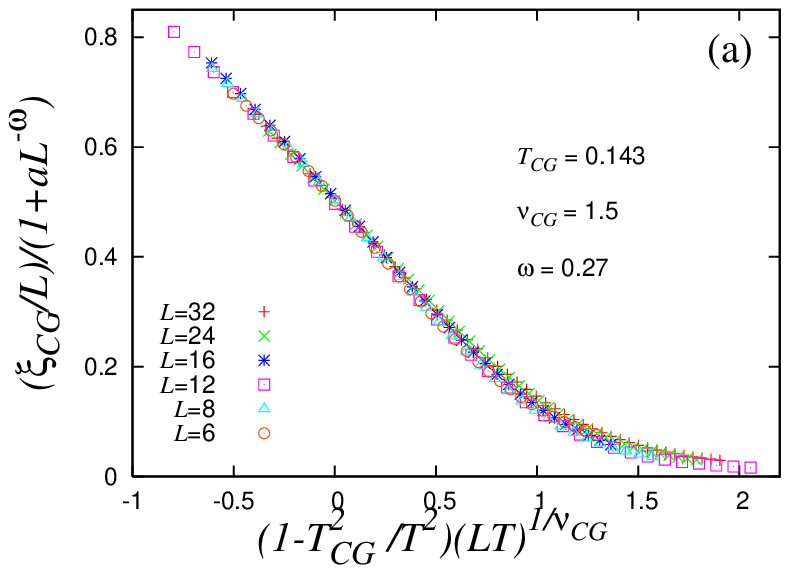}
\includegraphics[scale=0.9]{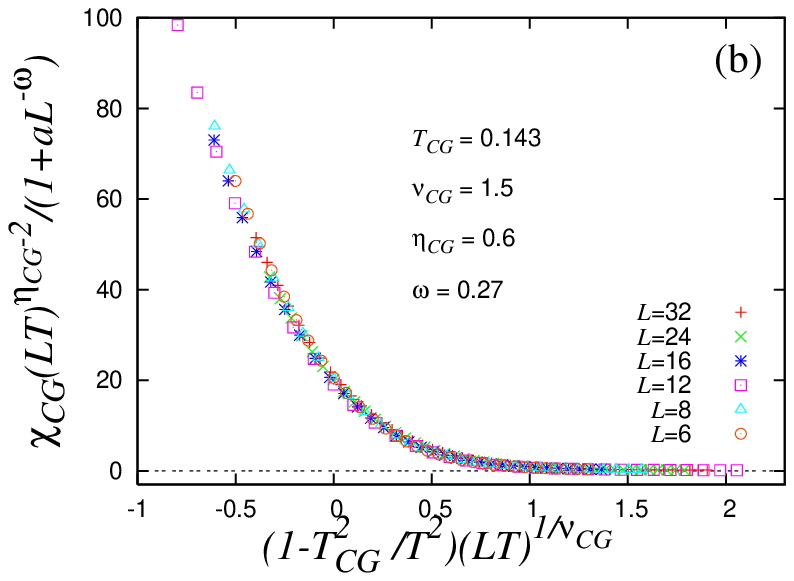}
\end{center}
\caption{
(Color online) Extended finite-size-scaling plots of the chiral-glass correlation-length ratio $\xi_{CG}/L$ (a), and of the chiral-glass susceptibility $\chi_{CG}$ (b) {\it \`a la\/} Campbell {\it et al\/} \cite{Campbell06}, where the correction-to-scaling effect is taken into account. The chiral-glass transition temperature and the leading correction-to-scaling exponents are fixed to $T_{CG}=0.143$ and $\omega+\frac{1}{\nu}=0.93$ as determined in \S 4. The best fit for $\xi_{CG}/L$ is obtained with $\nu_{CG}=1.5$, while that for $\chi_{CG}$ is obtained with $\nu_{CG}=1.5$ and $\eta_{CG}=0.6$.
}
\end{figure}

 Combining the exponent estimates obtained from $\xi_{CG}/L$ and $\chi_{CG}$, either by the standard analysis or by the extended one {\it \`a la\/} Campbell,  we finally quote as our best estimates of the CG exponents,
\begin{equation}
\nu_{CG}=1.4\pm 0.2\ ,\ \ \ \eta_{CG}=0.6\pm 0.2,
\end{equation}
while the correction-to-scaling exponent takes a rather small value, $\omega=\theta-\frac{1}{\nu}=0.3\pm 0.1$, suggesting that the correction-to-scaling effect is relatively large here. This value of $\omega$ is smaller than the corresponding value of the 3D Ising SG, $\omega \simeq 1$ \cite{Palassini99,Ballesteros00,Hasenbusch08}. 

 The estimated values of the CG critical exponents are compatible with the previous values obtained before for the same model $\nu_{CG} \simeq1.2$ and $\eta_{CG} \simeq 0.8$ \cite {HukuKawa00} and with those reported for the $\pm J$ 3D Heisenberg SG $\nu_{CG}=1.2(2)$ and $\eta_{CG}=0.8(2)$ \cite {HukuKawa05}. By contrast, the obtained CG exponents differ significantly from the standard exponent values of the 3D Ising SG, $\nu \simeq 2.5\sim 2.7$ and $\eta \simeq -0.38\sim -0.40$ \cite{Campbell06,Hasenbusch08}. The result unambiguously indicates that the chiral-glass transition belongs to a universality class distinct from that of the standard 3D Ising SG, although the underlying $Z_2$ symmetry is common between the two. Possible long-range and/or many-body nature of the chirality-chirality interaction might be the cause of this difference. Further study is required to clarify the cause of this difference.

 Our scaling analysis in this section were based on the CG correlation-length ratio and the CG susceptibility. One may wonder if what happens if one uses the CG Binder ratio in the analysis. As is already evident from the form of $g_{CG}$ shown in Fig.10(a), which exhibits a negative dip whose depth grows with the system size $L$, the finite-size scaling dose not work for $g_{CG}$ even with including the correction-to-scaling term. Such an exotic behavior of $g_{CG}$, {\it e.g.\/}, the existence of a growing negative dip and the non-monotonic size dependence observed in certain temperature range above $T_{CG}$, is most probably reflecting the peculiarity of the CG ordered state itself, a possible one-step-like RSB feature, not just the sub-leading correction-to-scaling effect. If the peculiar behavior of $g_{CG}$ arose reflecting the proximity of the nontrivial character of the CG ordered state, the finite-size-scaling analysis would not be applicable to $g_{CG}$ in a straightforward way, at least in the range of lattice sizes studied here. 

 Finally, we wish to refer to the critical properties of the SG transition which is deduced to occur at $T_{SG}\lsim 0.125$. In \S 4, we already estimated the critical-point-decay exponent from the size dependence of the SG order parameter $q_{SG}^{(2)}$ as $\eta_{SG}\lsim -0.30$. We also tried a finite-size-scaling analysis of both $\xi_{SG}/L$ and $\chi_{SG}$, either with or without a correction-to-scaling term, just as we performed for the CG transition. However, it turns out that the finite-size scaling analysis for the spin-related quantities does not work well for any choice of $T_{SG}$, $\nu_{SG}$ and $\eta_{SG}$, even if we adjust the assumed parameter values in a wide range. Possibly, the CG order that occurs preceding the SG order affects the scaling property of the spin-related quantities in a non-trivial way, and the data of still larger lattices and/or lower temperatures might be required to determine the critical properties of the SG transition.

\section{Summary}

 In summary, we have studied equilibrium ordering properties of the three-dimensional isotropic Heisenberg spin glass by means of extensive Monte Carlo simulations. By calculating various physical quantities including the correlation-length ratio, the Binder ratio, the glass order parameter and the overlap distribution function up to the size as large as $L=32$ and down to temperatures well below $T_{CG}$, we have given strong numerical evidence of successive CG and SG transitions occurring at $T_{CG}=0.143\pm 0.003$ and at $T_{SG}\leq 0.125^{+0.006}_{-0.012}$, respectively. The SG order sets in at a temperature at least about 10$\sim $15\% below the CG order, hence, the occurrence of the spin-chirality decoupling. On shorter length scale of $L\lsim 12$, the spin and the chirality often behave in a similar way, while, on longer length scale of $L\gsim 16$, the chirality shows a stronger ordering tendency than the spin. The observation supports the view of the trivial spin-chirality coupling at shorter length scale crossing over to the spin-chirality decoupling at longer length scale \cite{Kawamura07,Kawamura09}.

 One may feel that the relative distance between $T_{CG}$ and $T_{SG}$ is not so large, but, in fact, it is a sizable difference, much larger than the one observed in other systems exhibiting the spin-chirality decoupling, {\it e.g.\/}, the 2D regular frustrated {\it XY\/} model where the difference is known to be about 1\% \cite{OzekiIto,Hasenbusch05b,2DXYSG}. While the SG order in the 3D Heisenberg SG occurs at a nonzero temperature, as is consistent with the recent numerical works \cite{Matsubara00,Endoh01,Matsubara01,Nakamura02,LeeYoung03,BerthierYoung04,Picco05,Campos06,LeeYoung07}, it should be stressed that whether $T_{SG}$ is zero or nonzero is irrelevant to the chirality scenario of Refs.\cite{Kawamura92,Kawamura96,Kawamura07,Kawamura09} as long as the spin-chirality decoupling occurs, {\it i.e.\/}, $T_{SG} < T_{CG}$.

 We have observed a rather strong correction-to-scaling effect in our data, which, we have tried to control via the correction-to-scaling term with the correction-to-scaling exponent $\omega \simeq 0.3$. 
The analysis worked very well at least for the CG correlation length and the CG susceptibility. Hence, our conclusion of $T_{SG} < T_{CG}$ appears to be robust against the correction-to-scaling effect.

 We have also analyzed the critical properties associated with the CG transition. By mean of a finite-size scaling analysis with including the correction-to-scaling effect, we get an estimate of chiral-glass exponents $\nu_{CG}=1.4\pm 0.2$ and $\eta_{CG}=0.6\pm 0.2$. The possibility of a simultaneous spin and chiral transition of the KT-type as suggested in Refs.\cite{Campos06,LeeYoung07} is ruled out. We have shown that the behaviors of both the correlation-length ratio and the Binder ratio are entirely different from those of the 2D ferromagnetic {\it XY\/} model exhibiting the KT transition, and the KT scaling for these quantities does not work even with massive logarithmic corrections.

 The obtained values of the CG exponents are close to the values reported earlier in previous works, while they are entirely different from those of the 3D Ising SG. However, these CG exponents are impressively close to the experimental values of SG exponents of canonical SGs like CuMn, AuFe, AgMn, {\it etc\/}, {\it i.e.\/}, $\nu\simeq 1.3\sim 1.4$ and $\eta \simeq 0.5\sim 0.6$. Indeed, this coincidence gives a strong support to the chirality scenario of experimental SG transition of Ref.\cite{Kawamura92,Kawamura96,Kawamura07,Kawamura09}, since, in this scenario, the experimental SG exponents of weakly anisotropic Heisenberg-like SGs like canonical SGs are nothing but the CG exponents of the fully isotropic Heisenberg SG revealed via the random magnetic anisotropy. A very interesting consequence of the chirality scenario is that the chiral-glass transition, {\it not the spin-glass transition\/}, of the fully isotropic Heisenberg SG dictates the experimental SG transition. Experimentally, it remains highly interesting to directly estimate the set of chiral-glass exponents by means of high-precision Hall measurements \cite{TataraKawamura,Taniguchi04,Campbell04}. It might also be worthwhile to re-examine the standard spin-glass exponents for various Heisenberg-like SG materials by controlling the magnitude of magnetic anisotropy. 

  Although various physical quantities have consistently suggested that the SG order occurs at a temperature lower than the CG transition temperature, a precise estimate of the SG transition temperature and of the corresponding SG exponents still remains to be a rather difficult task, although we get $\eta_{SG} \lsim -0.30$. In any case, the critical properties of the SG transition definitely differ from those of the CG transition, since the associated $\eta$ values are largely different.

 By measuring the Binder ratio, the overlap distribution function and the non-self-averageness parameter $A$, we have observed that the chiral-glass ordered state is non-self-averaging and exhibits a nontrivial phase-space structure (RSB). More precisely, we have observed a strong similarity to the systems exhibiting the so-called one-step RSB. We note that the one-step RSB feature was also observed in the same model in an off-equilibrium simulation probing the breaking pattern of the fluctuation-dissipation relation \cite{Kawamura03}. According to the chirality scenario of Ref.\cite{Kawamura92,Kawamura96,Kawamura07,Kawamura09}, the properties of the SG ordered state of real canonical SGs should be governed by the properties of the CG ordered state of the fully isotropic Heisenberg SG. If so, one-step-like RSB should eventually be an attribute of the SG ordered state of real canonical SGs. This is in sharp contrast to the long-standing common belief in the community, {\it i.e.\/},  the SG ordered state of real canonical SGs exhibits either the hierarchical RSB (full RSB) or no RSB.

 After the submission of the manuscript, the authors learned that Fernadez {\it et al\/} also studied the same model by MC simulations up to the size $L=48$, and suggested that the spin and the chirality might order simultaneously \cite{Fernandez}. We wish to give a few comments here: First, we have confirmed that the data of $L\leq 32$ reported in Ref.\cite{Fernandez} now agree with our present data within the error bars. (This is somewhat in contrast to the data of Ref.\cite{LeeYoung07} which deviate from our present data by 5 to 6 of our $\sigma$ units.) Furthermore, the SG transition temperature reported in \cite{Fernandez} agrees with our present estimate $T_{SG}=0.125^{+0.006}_{-0.012}$. The major difference then concerns with the difference in the estimate of the chiral-glass transition temperature $T_{CG}$. The reason of this discrepancy seems to be primarily originated from their $L=48$ $\xi_{CG}/L$ data, which comes significantly smaller than the values expected from an extrapolation of the $L\leq 32$ data made in our present analysis. If the $L=48$ chiral data of Ref.\cite{VietKawamura} are to be trusted, it means a drastic changeover occurring in the chiral sector between the sizes $L=32$ and $L=48$. The physical origin of this size crossover, if any, has yet to be identified. Meanwhile, since we do not have at the moment a plausible explanation of such a size-crossover, and since the equilibration of $L=48$ chiral quantities is the hardest and the number of $L=48$ samples studied in Ref.\cite{Fernandez} (164 samples) is significantly smaller than that of other sizes (984 samples), we feel that the $L=48$ data of Ref.\cite{Fernandez} should be cross-checked carefully by independent calculations.

 Overall, we believe that our present data give strong numerical support to the view that the spin-chirality decoupling occurs in the 3D isotropic Heisenberg SG. Then, it also give support to the chirality scenario of experimental SG transitions.

\begin{acknowledgments}
The authors are thankful to I.A. Campbell, H. Yoshino and K. Hukushima for useful discussion. One of present authors (D.X.V.) also thanks to Mr. D. Tsuneishi and Mr. M. Nakamura for many helpful discussion. This study was supported by Grant-in-Aid for Scientific Research on Priority Areas ``Novel States of Matter Induced by Frustration'' (19052006). We thank ISSP, Tokyo University and YITP, Kyoto University for providing us with the CPU time.
\end{acknowledgments}

\begin{center}
    {\bf APPENDIX A: Simulation of the  ferromagnetic 2D {\it XY\/} model}
\end{center}

 In this appendix, we report on the results of our simulation on the ferromagnetic {\it XY\/} (plane rotator) model on a 2D square lattice,  a typical model exhibiting the KT transition. The Hamiltonian is given by
\begin{equation}
{\cal H}=-J\sum_{<ij>}\vec{S}_i\cdot \vec{S}_j\ \ ,
\label{eqn:hamil-2DXY}
\end{equation}
where $J>0$ is a ferromagnetic nearest-neighbor coupling and $\vec S_{i}$ is a two components classical unit vector at the site $i$. The lattice is a $L\times L$ square lattice ($L$ ranging from 8 to 512) with periodic boundary conditions.  We perform an equilibrium MC simulation by using the single-spin-flip Metropolis method and the over-relaxation method. The over-relaxation sweeps are repeated $M=L/8$ times per every Metropolis sweep, which constitutes our unit MC step.

The quantities we show here are the spin correlation-length ratio $\xi/L$ and the spin Binder ratio $g$. The spin correlation-length is defined by
\begin{equation}
\xi = 
\frac{1}{2\sin(k_\mathrm{m}/2)}
\sqrt{ \frac{ \langle m (\vec 0)^2 \rangle }
{\langle m (\vec{k}_\mathrm{m})^2 \rangle } -1 },
\end{equation}
where $\vec k = (k_{\mathrm{m}}, 0)$ with $k_{\mathrm{m}}=2\pi /L$, and $\langle\cdots\rangle$ denotes a thermal average, while
\begin{equation}
m(\vec k)^2 = \sum_{\mu =x,y}|\frac {1}{N} \sum_{i=1}^N S_{i\mu}\exp (i\vec k \cdot \vec r_{i})|^2
\end{equation}
is a $k$-dependent magnetization, $N$ being the total number of the spins. The spin Binder ratio is defined by
\begin{equation}
g=2-\frac {\langle m(\vec 0)^4 \rangle} {\langle m(\vec 0)^2 \rangle^2}.
\end{equation}
\begin{figure}[ht]
\begin{center}
\includegraphics[scale=0.9]{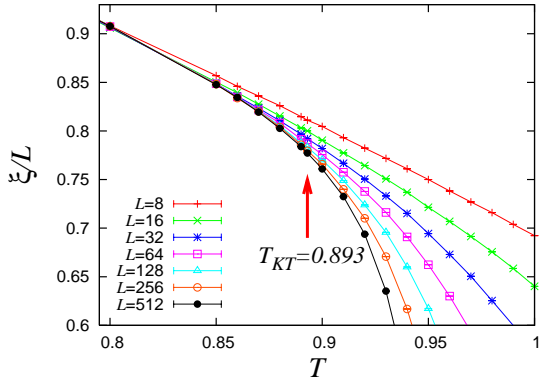}
\end{center}
\caption{
(Color online) The temperature and size dependence of the correlation-length ratio of the ferromagnetic 2D {\it XY\/} model. An arrow in the figure represents the location of the Kosterlitz-Thouless transition point.
}
\end{figure}

The temperature dependence of the correlation length ratio $\xi/L$ is shown in Fig.21. The KT transition temperature of this model was estimated rather precisely as $T_{KT}\simeq 0.893$ (in units of $J$) \cite {Hasenbusch05}. With increasing $L$, the $\xi/L$ curves do not cross at a finite temperature, but tend to merge progressively at temperatures lower than $T_{KT}$, as can be seen from the figure. One can see that the observed behavior of $\xi/L$ of the ferromagnetic 2D {\it XY\/} model is entirely different from the corresponding behavior of either the CG or SG correlation-length ratio, $\xi_{CG}/L$ or $\xi_{SG}/L$, of the 3D Heisenberg SG shown in Fig.9. The $\xi_{CG}/L$ and $\xi_{SG}/L$ curves of the 3D Heisenberg SG do not merge as in the $\xi/L$ curves of the 2D {\it XY\/} model but intersect, the crossing points shifting to lower temperatures for large $L$

\begin{figure}[ht]
\begin{center}
\includegraphics[scale=0.9]{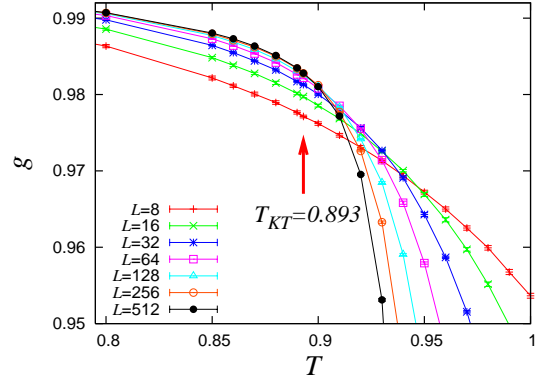}
\end{center}
\caption{
(Color online) The temperature and size dependence of the Binder ratio of the ferromagnetic 2D {\it XY\/} model. An arrow in the figure represents the location of the Kosterlitz-Thouless transition point.
}
\end{figure}

 Fig.22 exhibits the temperature dependence of the Binder ratio $g$ of the ferromagnetic 2D {\it XY\/} model. The $g$ curves of different $L$ now cross at a nonzero temperature. With increasing $L$, the crossing points approach $T_{KT}$ from above. The data for larger $L$ tend to give a ``merging'' behavior characteristic of the KT transition. This behavior is similar to the one reported by Loison for the same model \cite{Loison99b}. Again, one sees that the observed behavior of $g$ of the ferromagnetic 2D {\it XY\/} model is different from the corresponding behavior of either the CG or SG Binder ratio, $g_{CG}$ or $g_{SG}$, of the 3D Heisenberg SG shown in Fig.10. 

Hence, from the comparison of the correlation-length ratio $\xi/L$ and the Binder ratio $g$ of the ferromagnetic 2D {\it XY\/} model and of the 3D Heisenberg SG, one might also conclude that the transition of the 3D Heisenberg SG is not of the KT-type.

\begin{center}
    {\bf APPENDIX B: Simulations of the  ferromagnetic 3D $O(10)$ model}
\end{center}

 In this appendix, we report on the results of our simulations on the ferromagnetic $O(10)$ model on a 3D simple cubic lattice. The Hamiltonian is given by
\begin{equation}
{\cal H}=-J\sum_{<ij>}\vec{S}_i\cdot \vec{S}_j\ \ ,
\label{eqn:hamil-3DO10}
\end{equation}
where $J>0$ is a ferromagnetic nearest-neighbor coupling and $\vec S_{i}$ is a ten-components classical unit vector at the site $i$. In our simulation, we use the Me$_{\rm G}$ algorithm of Ref.\cite{Loison04} combined with the over-relaxation method. The lattice is a $L\times L\times L$ simple cubic lattice ($L$ ranging from 6 to 32) with periodic boundary conditions. The over-relaxation sweeps are repeated $M=L/2$ times per every Me$_{\rm G}$ sweep, which constitutes our unit MC step.

As in Appendix A, the quantities we show here are the spin correlation-length ratio $\xi/L$ and the spin Binder ratio $g$. The spin correlation-length ratio is defined by
\begin{equation}
\xi = 
\frac{1}{2\sin(k_\mathrm{m}/2)}
\sqrt{ \frac{ \langle m (\vec 0)^2 \rangle }
{\langle m (\vec{k}_\mathrm{m})^2 \rangle } -1 },
\end{equation}
where $\vec k = (k_{\mathrm{m}}, 0, 0)$ with $k_{\mathrm{m}}=2\pi /L$, and $\langle\cdots\rangle$ denotes a thermal average, while 
\begin{equation}
m(\vec k)^2 = \sum_{\mu =1}^{10}|\frac {1}{N} \sum_{i=1}^N S_{i\mu}\exp (i\vec k \cdot \vec r_{i})|^2
\end{equation}
is a $k$-dependent magnetization, $N$ being the total number of the spins. The spin Binder ratio is defined by
\begin{equation}
g=6-5\frac {\langle m(\vec 0)^4 \rangle} {\langle m(\vec 0)^2 \rangle^2}.
\end{equation}
\begin{figure}[ht]
\begin{center}
\includegraphics[scale=0.9]{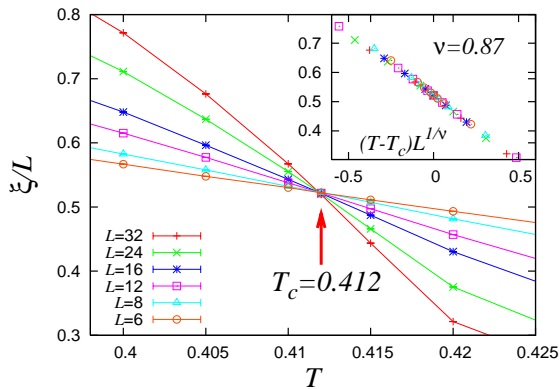}
\end{center}
\caption{
(Color online) The temperature and size dependence of the correlation-length ratio for the ferromagnetic 3D $O(10)$ model.  An arrow in the figure indicates the transition point. The inset exhibits the standard finite-size scaling plot without the correction term, where we put $T_c=0.412$ and $\nu =0.87$.
}
\end{figure}

The temperature dependence of the correlation length ratio $\xi/L$ is shown in Fig.23. As can be seen from the figure, the $\xi/L$ curves of various $L$ show a  clear crossing at an almost $L$-independent temperature $T_{c}=0.412\pm 0.002$ (in units of $J$), and splay out at lower temperatures.
\begin{figure}[ht]
\begin{center}
\includegraphics[scale=0.9]{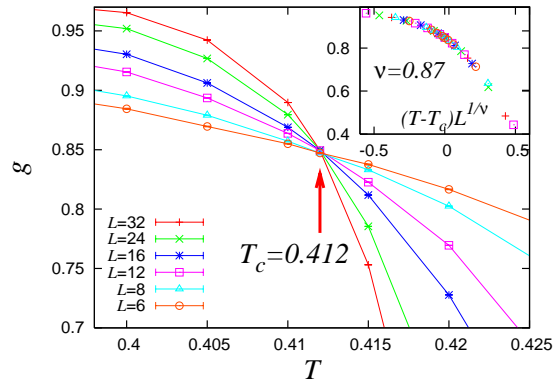}
\end{center}
\caption{
(Color online) The temperature and size dependence of the Binder ratio for the ferromagnetic 3D $O(10)$ model. An arrow in the figure indicates the transition point. The inset exhibits the standard finite-size scaling plot without the correction term, where we put $T_c=0.412$ and $\nu =0.87$.
}
\end{figure}

In Fig.24, we show the temperature dependence of the Binder ratio $g$ of the  ferromagnetic 3D $O(10)$ model.  As can be seen from the figure, the $g$ curves of different $L$ also show a very clear crossing at an almost $L$-independent temperature $T_{c}=0.412\pm 0.002$, and splay out  at lower temperatures. The behavior observed here for $g$ is quite similar to the one observed for $\xi/L$ in Fig.23. In particular, in spite of its large number of order-parameter component of $n=10$, no ``merging'' nor ``marginal'' behavior as suggested in Ref.\cite{PixleyYoung08} is observed. Very much similar ``crossing'' and ``splaying out'' behavior was observed also in the Binder ratio of the ferromagnetic 3D $O(6)$ model by Loison \cite{Loison99a}. 

 This observation clearly demonstrates that the peculiar ``non-crossing'' and ``negative dip'' behavior as observed in the spin Binder ratio $g_{SG}$ of the 3D Heisenberg SG shown in Fig.10(b) is not a trivial one originating from just the large number of order-parameter components ($n=9$ in the case of the Heisenberg SG). As discussed, the peculiar behavior observed in $g_{SG}$ of the 3D Heisenberg SG is likely to reflect an essential and peculiar feature of the ordered state of this model, most probably, the occurrence of a one-step-like RSB.
 
 Concomitantly, we also try to estimate the critical exponent $\nu$ from our data of $\xi/L$ and $g$ on the basis of the standard finite-size scaling analysis. Here, the correction seems to be negligible. Even without the correction term, we obtain a very good data collapse both for $\xi/L$ and $g$, as shown in the insets of Figs.23 and 24. We then get $\nu=0.87 \pm 0.03$.

%

%



\end{document}